\documentclass[aip,reprint]{revtex4-1}

\usepackage{lineno}
\usepackage{bm}
\usepackage{graphicx}
\usepackage{subfigure}
\usepackage{wrapfig}
\usepackage{epstopdf}
\usepackage{siunitx}
\usepackage{braket}
\usepackage{tabularx}
\usepackage{blindtext}
\usepackage{amsmath}
\usepackage{float}
\usepackage{booktabs,array}
\usepackage{booktabs}
\usepackage{tabularx}
\usepackage{multirow}
\usepackage{hyperref}
\usepackage[T1]{fontenc}
\usepackage{textcomp}
\usepackage{tabularx}
\usepackage{array}
\usepackage{colortbl}
\usepackage{wrapfig}
\usepackage{amsthm}
\usepackage{lipsum} 
\usepackage{color}
\usepackage{transparent}
\usepackage[section]{placeins}
\usepackage{afterpage}
\usepackage{amsmath}
\usepackage{subfiles}

\def\bem#1{\begin{mathletters}\label{#1}}
\def\eml{\end{mathletters}}

\def\4#1{{\boldsymbol{#1}}}
\def\8#1{{\widetilde{#1}}}

\usepackage{etoolbox}

\IfFileExists{Supplemantary_Notes.aux}{}{\immediate\write18{latexmk -pdf -f Supplemantary_Notes}}
\usepackage{xr}
\makeatletter

\def\@email#1#2{%
 \endgroup
 \patchcmd{\titleblock@produce}
  {\frontmatter@RRAPformat}
  {\frontmatter@RRAPformat{\produce@RRAP{*#1\href{mailto:#2}{#2}}}\frontmatter@RRAPformat}
  {}{}
}
\def\frontmatter@RRAPformat{\centering}
\def\frontmatter@authorformat{\centering}
\def\frontmatter@affiliationformat{\centering}

\patchcmd{\frontmatter@title@produce}
{\@author}
{\vspace{1.5em}\@author}
{}{}

\makeatother
\begin{document}
\renewcommand\linenumberfont{\normalfont\tiny}
\title{Biocompatible Vaterite Carriers Enable Multimodal Quantum Sensing with Nanodiamonds.}
\author{T. Amro}
\affiliation{Institute of Applied Physics, The Hebrew University, Jerusalem 91904, Israel}
\affiliation{Triangle Regional Research and Development Center, Kfar Qara’ 3007500, Israel}

\author{M. Attrash}
\affiliation{Triangle Regional Research and Development Center, Kfar Qara’ 3007500, Israel}
\affiliation{School of Electrical Engineering, Tel Aviv University, Ramat Aviv, Tel Aviv 69978, Israel}

\author{A. Droby}
\affiliation{Physics Program, Graduate Center, The City University of New York, New York, NY 10016, US}

\author{A. Ushkov}
\affiliation{School of Electrical Engineering, Tel Aviv University, Ramat Aviv, Tel Aviv 69978, Israel}

\author{R. Malkinson}
\affiliation{Institute of Applied Physics, The Hebrew University, Jerusalem 91904, Israel}

\author{P. Penshin}
\affiliation{Institute of Applied Physics, The Hebrew University, Jerusalem 91904, Israel}

\author{A. Hen}
\affiliation{Institute of Applied Physics, The Hebrew University, Jerusalem 91904, Israel}

\author{V. Bobrovs}
\affiliation{Institute of Telecommunications, Riga Technical University, Riga LV-1048, Latvia}

\author{P. Ginzburg}
\affiliation{School of Electrical Engineering, Tel Aviv University, Ramat Aviv, Tel Aviv 69978, Israel}

\author{H. Barhum$^*$}
\affiliation{Triangle Regional Research and Development Center, Kfar Qara’ 3007500, Israel}
\affiliation{School of Electrical Engineering, Tel Aviv University, Ramat Aviv, Tel Aviv 69978, Israel}

\author{N. Bar-Gill$^*$}
\affiliation{Institute of Applied Physics, The Hebrew University, Jerusalem 91904, Israel}
\affiliation{The Racah Institute of Physics, The Hebrew University, Jerusalem 91904, Israel}
\affiliation{The Center for Nanoscience and Nanotechnology, The Hebrew University of Jerusalem, Jerusalem 91904, Israel}
\date{\today} 
\email{tamara.amro@mail.huji.ac.il}
\begin{abstract}
Mobile nanodiamond quantum sensors in liquids are limited by Brownian rotation, variable photon collection and perturbations from optical trapping. Here we assemble 40-nm nitrogen-vacancy nanodiamonds on porous, birefringent vaterite microspherulites, creating sensors with a polarization-addressable body frame and a chemically active carbonate interface. Under 976-nm trapping, the sensors retain spin resonance and longitudinal relaxation, with less than 7\% contrast variation and an approximately 1-MHz resonance shift at 0.8~W. Zeeman-split resonances resolve magnetic fields from 0 to 0.8~mT, with a response metric of 78--144~$\mu$T~Hz$^{-1/2}$. In cell-culture medium, a 10.7-$\mu$M proton-equivalent dose shortens $T_1$ from $23.4 \pm 2.3$ to $9.0 \pm 1.2~\mu$s, yielding concentration and pH sensitivities of 6.46~$\mu$M~Hz$^{-1/2}$ and 6.54~mPH~Hz$^{-1/2}$. A 500-fold larger proton dose in ethanol produces a weaker response. A grand-canonical charge-regulation model links proton chemical potential to interfacial switching, establishing a route to multimodal quantum sensing in complex liquids.
\end{abstract}
\vspace{0.8em}
\maketitle

\begin{figure*}[t]
    \centering
    \includegraphics[width=\linewidth]{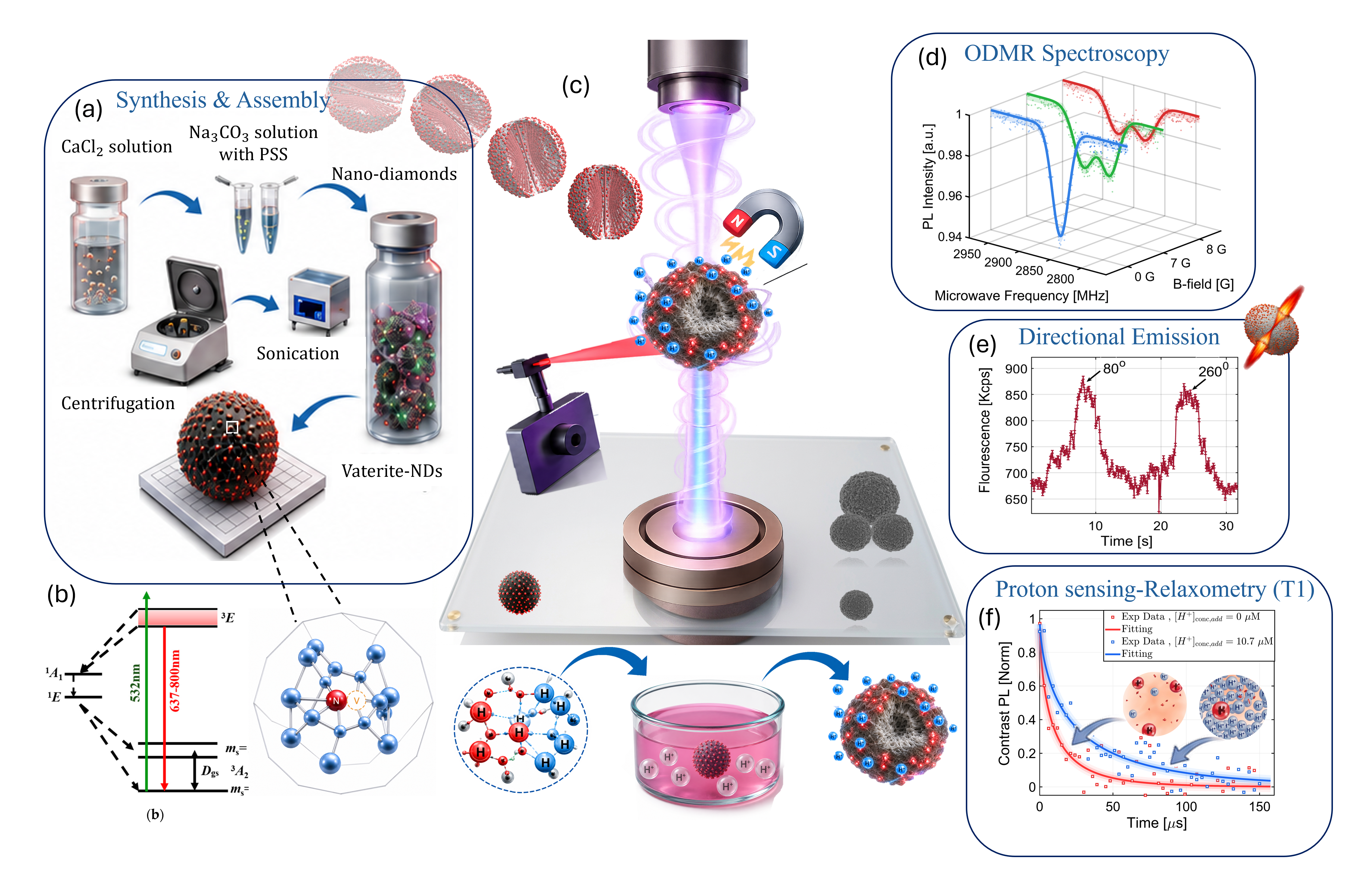}
    \caption{\textbf{Carrier-engineered vaterite--nanodiamond quantum sensor}
    (a) Assembly of fluorescent 40-nm NV nanodiamonds on vaterite microspherulites.
    (b) NV$^-$ centre energy levels and spin-dependent optical cycle.
    (c) Optical trapping and confocal readout of one hybrid in liquid.
    (d) ODMR spectra under applied magnetic fields.
    (e) Polarisation-dependent photoluminescence from the birefringent carrier.
    (f) $T_1$ relaxometry during nominal proton-equivalent dosing in buffered medium.
    }
    \label{fig:1}
\end{figure*}
Nitrogen-vacancy (NV) centers in diamond combine optical spin initialization and readout with room-temperature sensitivity to magnetic fields, surface noise and chemical dynamics\cite{ref1,ref2,ref3,ref4,ref5,ref6,ref7,ref8,ref9,ref10,ref11}. Nanodiamonds extend these functions to dispersible probes. In a liquid the sensor itself becomes part of the measurement. Brownian translation changes excitation and collection. Brownian rotation changes the NV–field projection. Near-infrared trapping can also perturb charge and spin dynamics \cite{ref12,ref13,ref14}. The central challenge is therefore to make a moving particle behave as a reproducible quantum probe without separating its active surface from the liquid. We address this problem by engineering the mesoscale carrier rather than the diamond alone.40 nm NV nanodiamonds are assembled at the
surface of porous 3--5-\ensuremath{\mu\mathrm{m}} vaterite microspherulites.Vaterite contributes optical birefringence and a carbonate-rich interface \cite{ref15,ref16}. The nanodiamonds retain the spin-triplet transducer
\cite{ref17}.The anisotropic carrier supports non-Mie resonances \cite{ref18}.Longitudinal relaxation samples the interfacial spectral density \cite{ref19}.The micrometre-scale hydrodynamic volume suppresses angular diffusion relative to an isolated 40-nm nanodiamond. Birefringent optical torque adds an orienting potential. This hierarchy supplies an addressable body frame without withdrawing the NV layer from the liquid.It reduces the orientation and linewidth variations associated with freely moving nanodiamonds \cite{ref20,ref21}.Fig.\ref{fig:1} maps the resulting
sequence from assembly to optical, Zeeman, and chemical readout. 

\section{Surface-localised quantum layer}
The carrier is formed by polyelectrolyte-assisted assembly of
fluorescent nanodiamonds onto vaterite microspherulites (Fig.\ref{fig:1} a).The
3--5-\ensuremath{\mu\mathrm{m}} host is sufficiently large to produce an experimentally
resolvable polarisation response under optical torque.The 40-nm
nanodiamonds preserve a high surface-to-volume ratio and place a
substantial fraction of NV centres near a liquid-accessible boundary.This division of length scales is central to the design. Vaterite defines the body frame and photonic environment; the nanodiamonds
provide spin-dependent fluorescence; the vaterite/poly(sodium
4-styrenesulfonate) (PSS)/nanodiamond junction supplies the chemically responsive interface.

\begin{figure*}[t]
  \centering
\includegraphics[width=\linewidth]{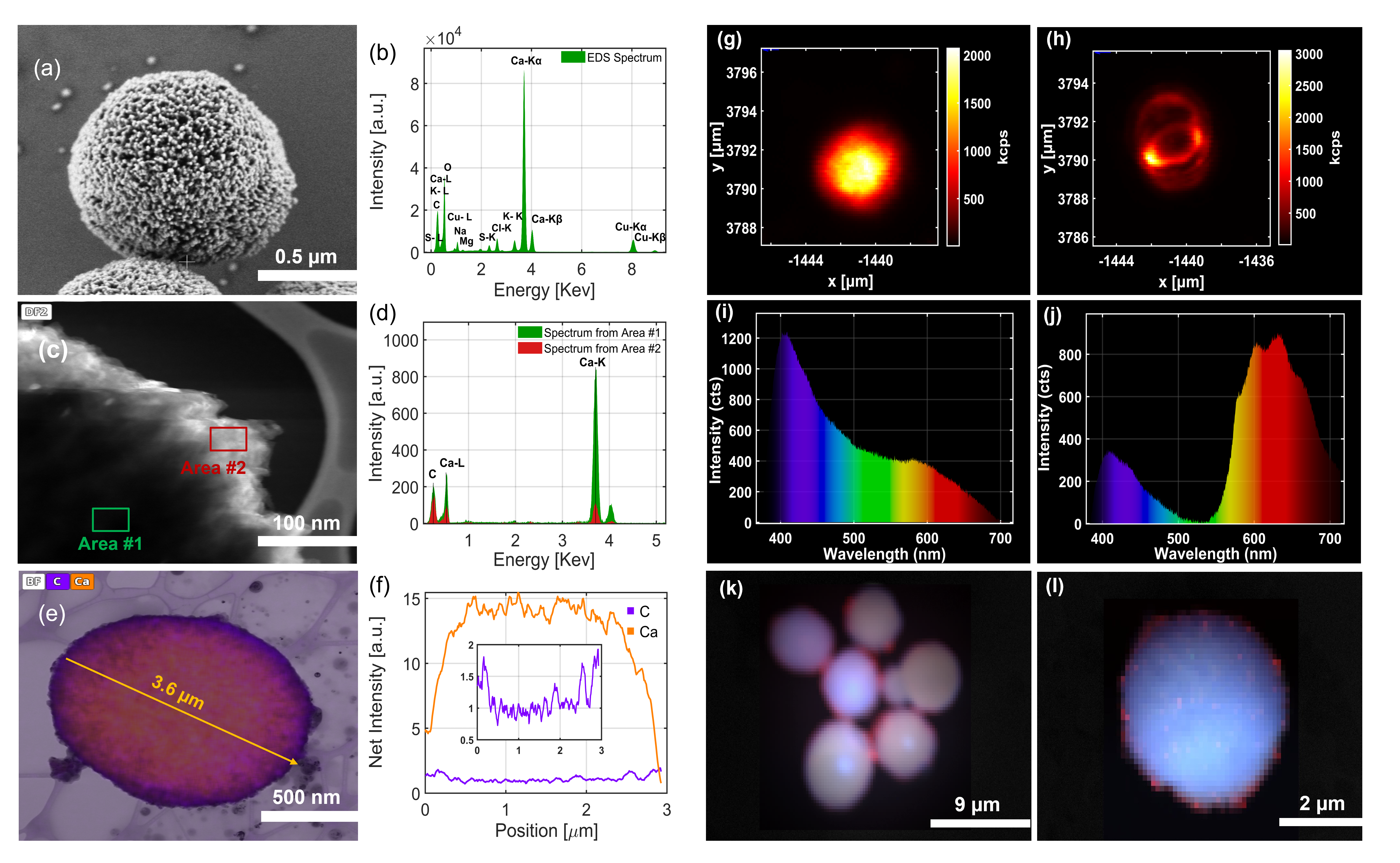}
\caption{\textbf{The NV layer is localised at the carrier--liquid interface.}
(a) SEM image of a representative hybrid particle.
(b) Survey EDS spectrum showing the calcium-rich host.
(c) TEM image of the particle edge.
(d) Local EDS spectra from the marked interior and peripheral regions.
(e) Bright-field TEM image with carbon (purple) and calcium (orange) overlays for a 3--3.6~$\mu$m particle.
(f) Elemental line profiles across the particle.
(g,h) Confocal PL maps acquired at the lower surface and mid-plane.
(i,j) CL spectra from the carrier core and surface region.
(k,l) CL maps acquired with a 30-keV electron beam at 17 and 16 nA.
The combined structural and spectral data are consistent with a predominantly surface-bound NV$^{-}$ nanodiamond layer.}
\label{fig:fig2}
\end{figure*}

The structural and optical data in Fig.~\ref{fig:fig2} are read from carrier morphology to quantum-emitter localisation. Scanning electron microscopy shows a nearly spherical surface that is rough and porous (Fig.~\ref{fig:fig2} a). Survey energy-dispersive X-ray spectroscopy identifies calcium as the dominant host element (Fig.~\ref{fig:fig2} b). Transmission electron microscopy resolves nanoscale material at the boundary (Fig.~\ref{fig:fig2} c).Local spectra distinguish the calcium-rich interior from the carbon-rich periphery (Fig.~\ref{fig:fig2} d).The carbon/calcium overlay and line profile show a calcium-dominated core with a carbon-enriched outer region extending over approximately 150--200 nm (Fig.~\ref{fig:fig2} e,f). PSS also contains carbon. Elemental mapping alone therefore does not uniquely identify nanodiamond; it establishes a surface-decorated carbonaceous layer that must be cross-checked optically.
 \\
Confocal photoluminescence (PL) provides that cross-check.A scan of the lower surface reveals bright NV emission across the particle footprint (Fig.~\ref{fig:fig2} g).A mid-plane scan instead produces a peripheral ring with weak signal from the interior (Fig.~\ref{fig:fig2} h). Cathodoluminescence (CL) separates the host and sensor spectra. The vaterite core exhibits a blue violet band near 405 nm (Fig.~\ref{fig:fig2} i).The surface region contains an additional red band centred near 632 nm within the NV emission window (Fig.~\ref{fig:fig2} j). The corresponding maps retain the same spatial separation (Fig.~\ref{fig:fig2} k,l).Together, the electron microscopy, elemental contrast and spectrally resolved luminescence place the optically active NV layer
predominantly at the carrier-liquid interface.
\\This geometry does more than expose the sensor chemically.It positions the emitters at the curved, anisotropic boundary, where the optical momentum depends on carrier orientation.The local density of states and far-field collection also depend on that orientation.Figure ~\ref{fig:fig2} therefore establishes the surface-localised NV layer; Figure 3 next tests how that geometry reshapes emission and optical torque.


\section{Birefringence defines the optical frame}
\begin{figure}[h]
\centering
\includegraphics[width=0.5\textwidth]{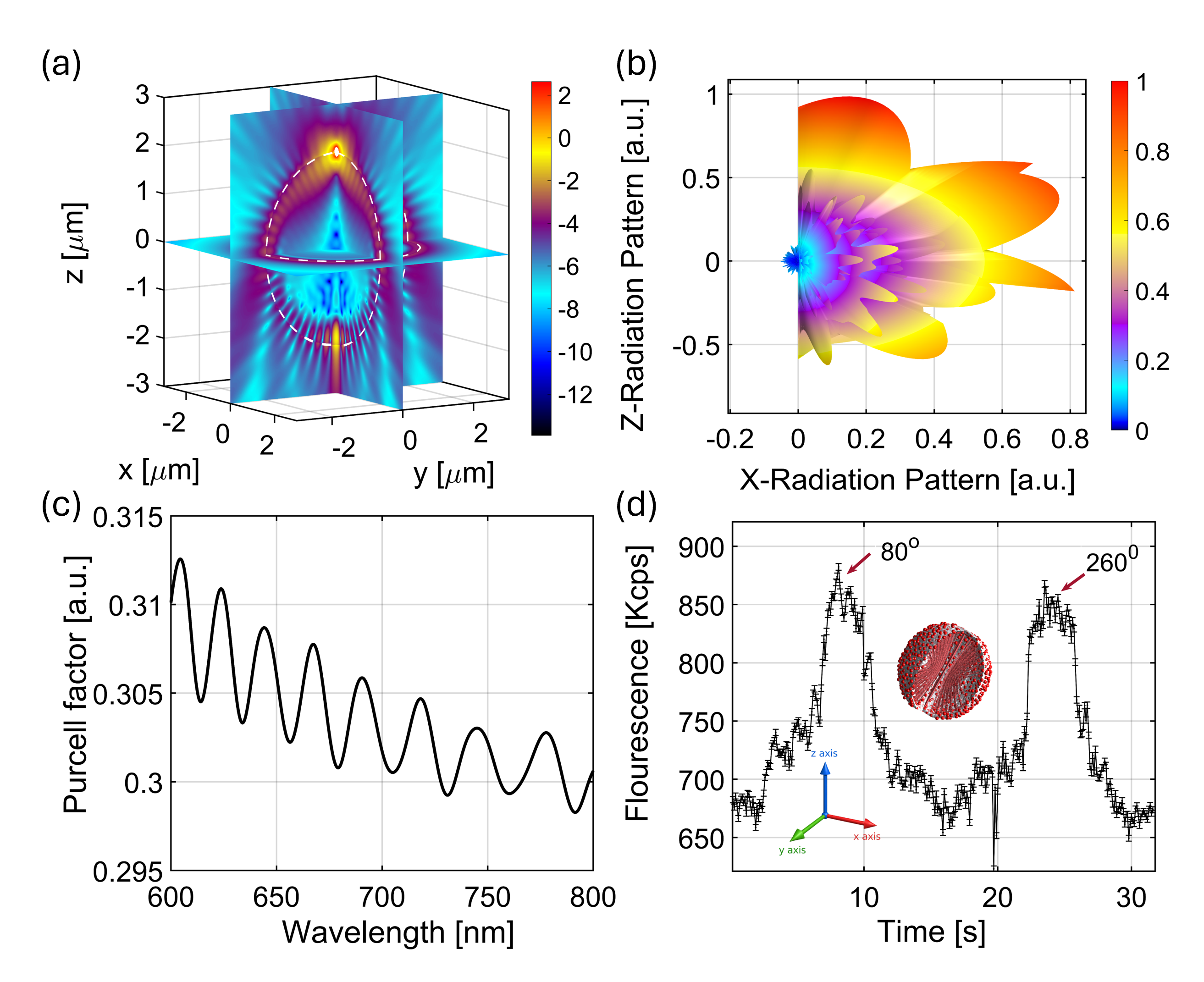}
\caption{\textbf{Birefringence couples carrier orientation to emission collection.} (a) Calculated near-field distribution at 680 nm for a z-oriented dipole in a 40-nm nanodiamond at the vaterite surface. (b) Corresponding far-field distribution at 650 nm. (c) Purcell-factor spectrum averaged over three dipole orientations and multiple surface positions relative to an identical emitter in homogeneous water. Values below unity indicate suppression of total radiated power relative to that reference, while the angular distribution remains anisotropic.(d) PL from one hybrid trapped in D\ensuremath{_{2}}O, averaged for 20 s at each quarter-wave-plate setting. Two maxima separated by approximately 180\ensuremath{^{\circ}} report a reproducible polarisation-dependent body axis; three-particle data are in Supplementary Fig. \ref{fig:B1.1},\ref{fig:B1.2},\ref{fig:B2}.}
\label{fig:fig4}
\end{figure}

Vaterite is an anisotropic dielectric. In a linearly polarised trap its ordinary and extraordinary field components experience different refractive indices. This produces an optical torque that tends to align a body-fixed axis with the incident polarisation.To lowest order the orienting energy is \ensuremath{U(\theta)\simeq-\Delta\alpha |E|^2\cos^2\theta/4} and \ensuremath{\tau_\theta=-\partial U/\partial\theta}. Its ratio to \ensuremath{k_\mathrm{B}T} sets the competition with Brownian rotation \cite{ref33,ref34}.Surface-bound NV dipoles simultaneously couple to the curved dielectric boundary. We modelled these linked optomechanical and photonic effects with finite-difference time-domain simulations of a 40-nm diamond sphere containing an electric dipole at the surface of a birefringent vaterite sphere in water (Supplementary Note B). Figure ~\ref{fig:fig4} shows the optical response from the local electromagnetic
field to the measured signal. A surface dipole aligned with the model optical axis develops a forward near-field lobe (Fig. \ref{fig:fig4} a). The transformed far field becomes strongly asymmetric (Fig. \ref{fig:fig4} b).Averaging
over dipole orientations and surface positions gives a Purcell factor near 0.30 relative to homogeneous water. It remains below unity across the simulated NV band (Fig. \ref{fig:fig4} c). The calculation therefore predicts angular redistribution rather than a net spontaneous-emission-rate enhancement. In Green's function language the anisotropic boundary redirects emission into the collection cone \cite{Lukosz:79}. The corresponding experiment records PL from a trapped hybrid as the quarter-wave-plate setting is varied. This produces the polarisation-dependent response in Fig. \ref{fig:fig4} d.
\\
Single hybrids were trapped in D\ensuremath{_{2}}O with a 975--976-nm trapping beam. Rotating the quarter-wave plate changed the trapping polarisation state and the optical torque applied to the birefringent carrier. The detected PL varied periodically. Two maxima were separated by approximately 180\ensuremath{^{\circ}} (Fig.\ref{fig:fig4} d and Supplementary Fig.\ref{fig:B2}). This reproducible angular response is consistent with an optically addressable carrier orientation and
orientation-dependent collection.The particle angle was not independently tracked in these measurements. We therefore describe the state as orientation-stabilised rather than quantitatively
orientation-locked.
\section{Magnetic sensing with stable ODMR in liquid}
\begin{figure}[h]
\centering
\includegraphics[width=0.48\textwidth]{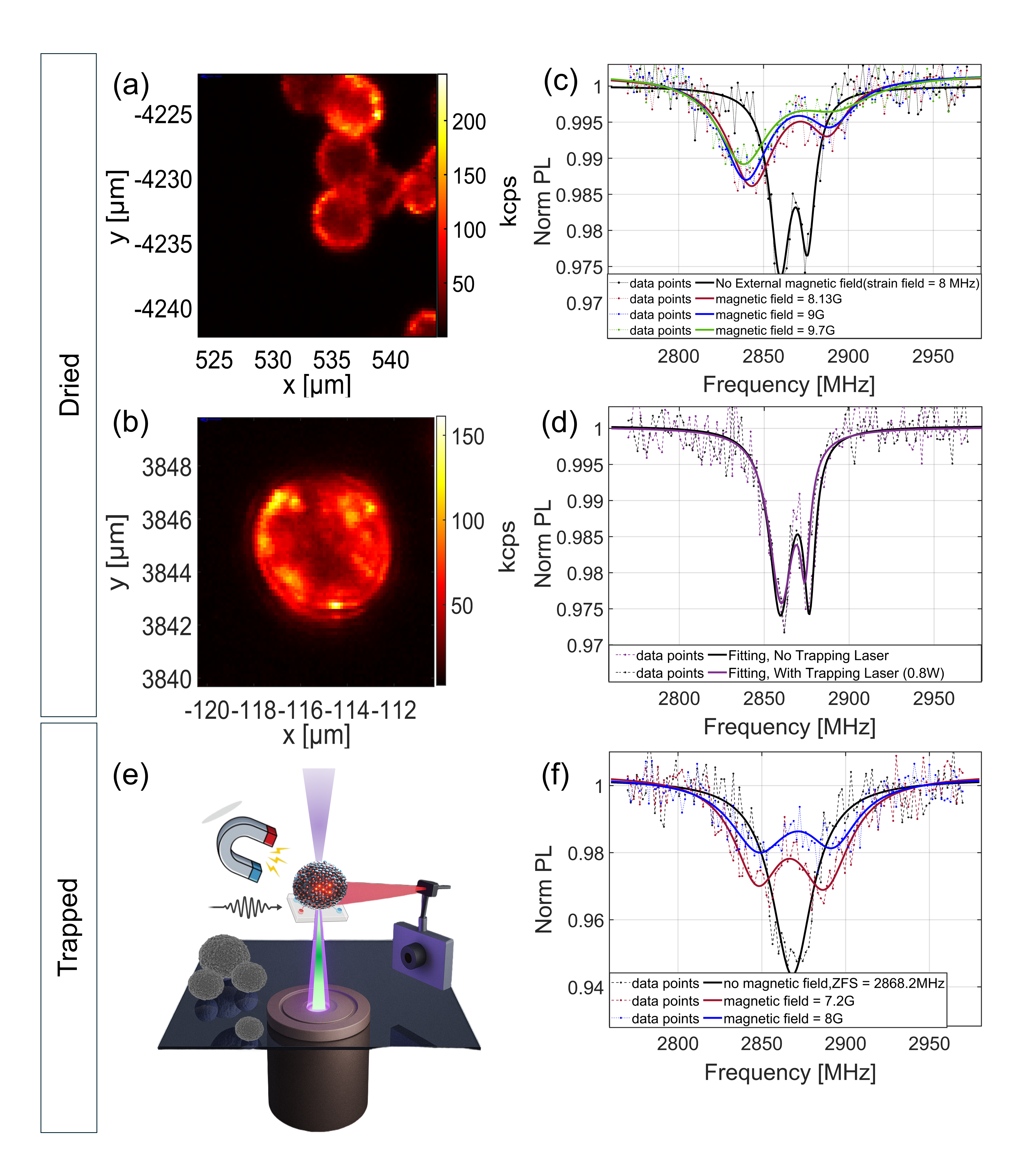}
\caption{\textbf{ODMR of vaterite-nanodiamond hybrid microcarriers.} (a) Wide-field PL image of multiple hybrids.(b) PL image of a single hybrid selected for spectroscopy. (c) ODMR spectra of a dried hybrid under different magnetic fields, showing Zeeman splitting around ($D \approx 2868.2$)\,MHz.(d) ODMR spectra of a dried hybrid acquired with 976\,nm illumination from the trapping path switched off and on, showing negligible change in linewidth or resonance position.(e) Schematic of the optical trapping and detection setup.(f) ODMR spectra of a trapped hybrid at different magnetic fields; the spectra were fit with two Lorentzian dips for sensitivity estimation; the corresponding magnetic sensitivity is 171.7\,\textmu T/$\sqrt{\text{Hz}}$.}
\label{fig:fig5}
\end{figure}

The negatively charged NV centre is widely used as a nanoscale magnetic-field sensor, owing to its optically addressable, spin-triplet ground state.This state is described in frequency units by equation \ref{eq:main-1}.Here D and E are in hertz.
The electron gyromagnetic ratio \ensuremath{\gamma}\ensuremath{_{e}} is in rad s\ensuremath{^{-1}} T\ensuremath{^{-1}}. B is in tesla and
the omitted terms collect additional local interactions \cite{ref17}. Static
fields lift the \ensuremath{m_s} = \ensuremath{\pm}1 degeneracy through the Zeeman interaction.Transverse strain and electric fields contribute to E. In freely diffusing nanodiamonds, Brownian rotation continuously reorients the NV axis relative to the laboratory field, converting this motion into frequency noise and together with fluctuating collection geometry, degrading ODMR contrast\cite{ref20,ref21}. The vaterite orienting potential suppresses this variation but not static ensemble broadening. Figure \ref{fig:fig5} examines whether carrier-level optical control remains compatible with this Hamiltonian, proceeding from dry calibration and the 976 nm perturbation test to a single hybrid particle trapped in liquid.
\begin{equation}
\frac{\widehat{\mathcal H}}{h}
=D\widehat S_z^2
+E\left(\widehat S_x^2-\widehat S_y^2\right)
+\frac{\gamma_e}{2\pi}\mathbf B\cdot\widehat{\mathbf S}
+\frac{\widehat{\mathcal H}_{\mathrm{hf}}}{h}
+\cdots
\label{eq:main-1}
\end{equation}
ODMR spectra acquired on dried hybrids reveal the characteristic NV$^-$ resonance at 2.867--2.868 GHz. The
approximately 8-MHz transverse splitting is consistent with enhanced local strain from the small nanodiamond size and residual surface strain introduced during vaterite shell assembly (Fig.\ref{fig:fig5} c). Applied fields produce resolved Zeeman splitting and confirm that spin addressability is retained after assembly.Introducing 976-nm illumination through the trapping path leaves the linewidth unchanged within fit uncertainty.Contrast varies by less than 7\%.The fitted resonance centre changes by approximately 1 MHz — approximately 0.035$\%$ of the ~2868 MHz zero-field splitting — between the trap-off and 0.8 W trap-on conditions (Fig. \ref{fig:fig5} d). This compares favourably with a resonance shift of ~1.58  MHz reported at 60 mW for 1064 nm trapping of FNDs in water \cite{ref10}, where photothermal heating and NIR-driven charge-state redistribution were identified as the dominant perturbation mechanisms, despite our 976 nm beam operating at more than an order of magnitude higher trapping power. The substantially reduced relative drift is consistent with 976 nm lying farther from the NV photoionisation band, and accords with prior comparative studies of trapping wavelength and NV coherence \cite{ref13,ref14}.
A single hybrid trapped in ethanol retains resolved Zeeman-split resonances from 0 to 0.8 mT (0--8 G) with contrasts of 2--8\% (Fig. \ref{fig:fig5} f  and Supplementary Fig. \ref{fig:C1}). Under the source-analysis convention, Lorentzian fits give a reported field-response metric of 78--144 \ensuremath{\mu\mathrm{T}} Hz\ensuremath{^{-1/2}}. The measurement establishes liquid-phase ODMR from the trapped microcarrier. A fully noise-equivalent sensitivity requires the acquisition time together with photon-counting variance and the independent-particle distribution.
Broad ensemble linewidths and modest contrast limit the response. Heterogeneity in nanodiamond orientation and local strain adds a further limit. Larger nanodiamonds have reached approximately 50 \ensuremath{\mu\mathrm{T}}
Hz\ensuremath{^{-1/2}} under related trapping conditions \cite{ref22}.Pulsed ODMR plus time-gated detection and temporal filtering offer established routes to improve photon and spin-readout efficiency \cite{ref23,ref24}. The achievable
gain for this carrier must be measured rather than extrapolated. We estimate that implementing such optimizations could enhance sensitivity by approximately two orders of magnitude, bringing performance to the level of the current state-of-the-art.
\section{Interfacial relaxometry and proton‑noise transduction}
\begin{figure*}[t]
  \centering
   \includegraphics[width=\linewidth]{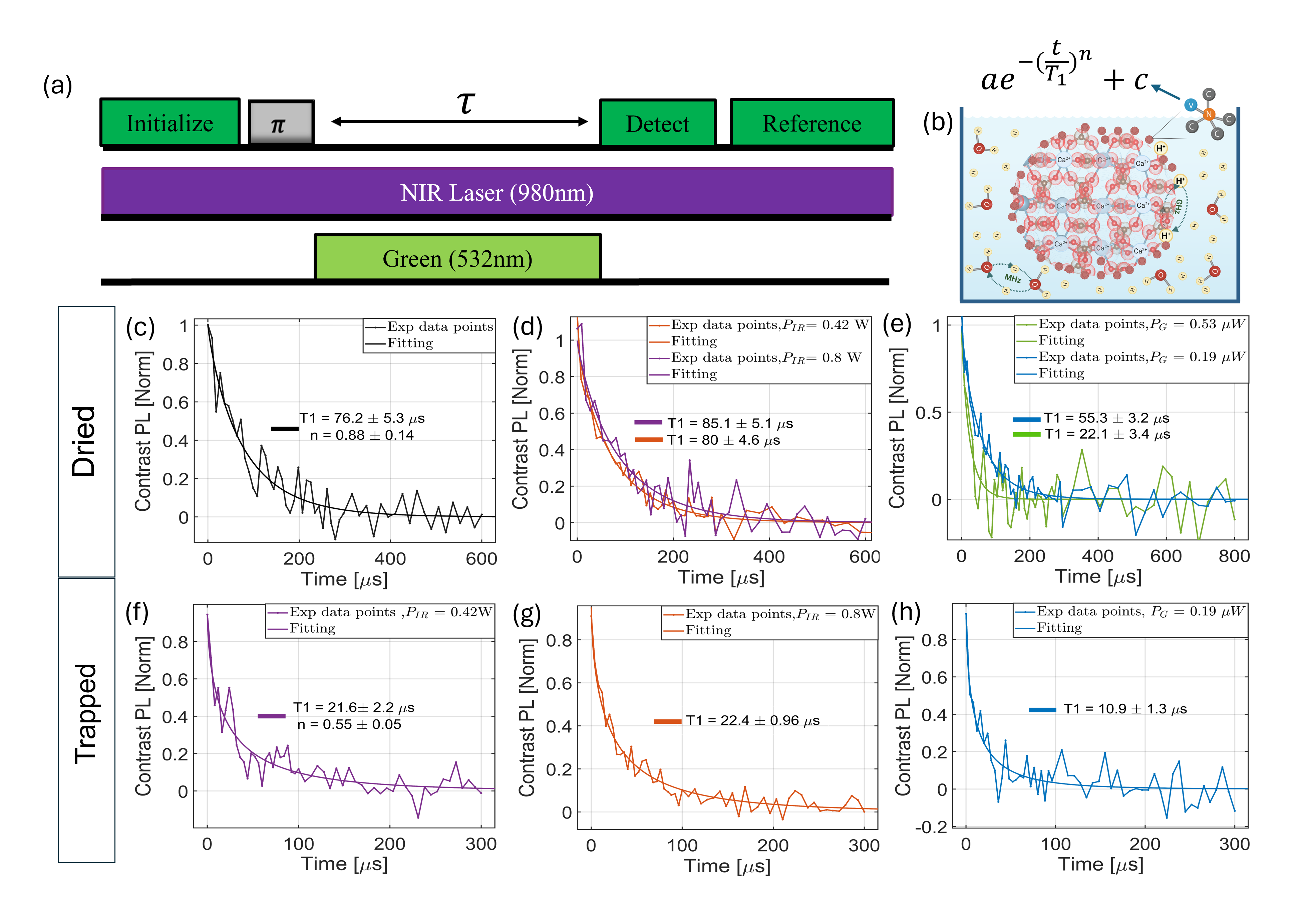}
    \caption{\textbf{Optical controls separate trapping from interfacial relaxation.} (a) Pulse sequences comprising optical initialisation, a variable dark interval $\tau$, optical readout and a reference measurement, with a microwave $\pi$ pulse applied for the dried-state measurements but omitted for the trapped-state measurements. The 976-nm trap remains on; 532-nm light is absent during the dark interval unless tested.(b) Interfacial fluctuations that can drive NV relaxation. (c) \ensuremath{T_1} decay of a dried hybrid. (d) Dried-hybrid \ensuremath{T_1} under 976-nm illumination. (e) Dried-hybrid \ensuremath{T_1} under weak 532-nm illumination during the relaxation interval. (f,g) \ensuremath{T_1} decays of trapped hybrids in ethanol containing up to 10\% water. (h) \ensuremath{T_1} under simultaneous 532- and 976-nm illumination. Each displayed decay is from the selected hybrid for that acquisition; curves are stretched-exponential fits and quoted uncertainties are nonlinear-fit uncertainties.}
  \label{fig:5}
\end{figure*}

\begin{figure*}[t]
\centering
\includegraphics[width=\linewidth,height= 1.2\textheight,keepaspectratio]{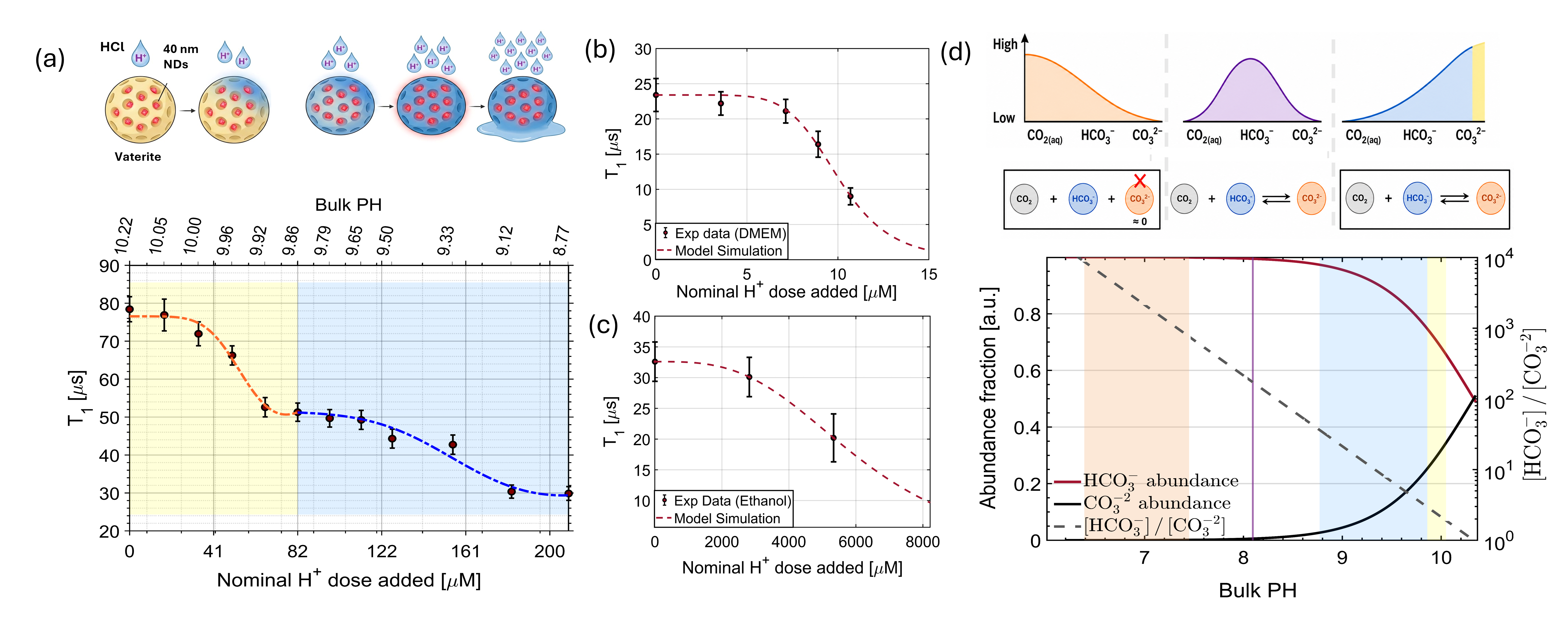}
\caption{\textbf{Proton-regulated relaxation depends on the surrounding medium.} (a) NaOH/HCl titration; the lower axis reports nominal H\ensuremath{^{+}} dose and the upper axis measured bulk pH.(b) \ensuremath{T_1} in open-air DMEM during addition of pH 4.01 potassium hydrogen phthalate buffer. Because bulk pH remains 8.100–8.088, the x axis is nominal proton-equivalent dose.(c) Ethanol control under much larger nominal additions. (d) Aqueous carbonate-pair coordinate {[}HCO\ensuremath{_3^{-}}{]}/{[}CO\ensuremath{_3^{2-}}{]}. Points in b and c are fit-derived \ensuremath{T_1} values from the selected hybrid series; error bars are nonlinear-fit uncertainties. Dashed curves are empirical regional or hill-like projections, not global fits of equations \ref{eq:main-3}-\ref{eq:main-6}.}
\label{fig:6}
\end{figure*}

Longitudinal relaxation provides an indirect chemical-environment readout channel.ODMR reports resonance-frequency shifts. \ensuremath{T_1} samples environmental
fluctuations with spectral weight at the allowed NV transitions.The NV
therefore acts as a frequency-selective noise filter rather than a
direct counter of nearby protons. Equation (\ref{eq:main-2}) separates the fitted
comparison metric \ensuremath{\Gamma_{1,\mathrm{eff}}} \ensuremath{\equiv} 1/\ensuremath{T_1} from one transverse-magnetic Bloch--Redfield convention \cite{ref3,ref4,ref6,ref19,ref25}.In that convention \ensuremath{S_{B_i}(\omega)}
is an angular-frequency field power spectral density in T\textsuperscript{2}.s and \ensuremath{\gamma}\ensuremath{_{e}} is
in rad s\ensuremath{^{-1}} T\ensuremath{^{-1}}. The ensemble prefactor absorbs PSD normalisation and
$NV^{-}$orientation weighting. Several non-magnetic processes can regulate the measured decay. We therefore describe a fast relaxation-active interfacial bath rather than direct proton-spin detection.
\begin{equation}
\begin{aligned}
\Gamma_{1,\mathrm{eff}}&\equiv T_1^{-1},\\
\Gamma_1^{(\perp B)}&=\Gamma_{1,0}+\frac{\gamma_e^2}{2}
\left[S_{B_x}(\omega_0)+S_{B_y}(\omega_0)\right].
\end{aligned}\label{eq:main-2}
\end{equation}
Fig. \ref{fig:5} separates the optical contribution from the mechanical and environmental contributions in measurement order.Panels a and b define the pulse sequence and interfacial-noise concept. A dried hybrid gives
the baseline decay (Fig. \ref{fig:5} c).A 976-nm power series leaves \ensuremath{T_1} in the 76--80-\ensuremath{\mu\mathrm{s}} range within uncertainty (Fig. \ref{fig:5} d). Weak 532-nm illumination
during the dark interval shortens \ensuremath{T_1} by about 70\% to approximately 22
\ensuremath{\mu\mathrm{s}} (Fig. \ref{fig:5} e). Infrared confinement is therefore compatible with the
reported sequence. Green excitation must remain restricted to preparation and readout unless tested explicitly.
\\Immersion activates the interfacial bath. Ethanol containing up to
10\% water \ensuremath{T_1} shortens to approximately 21--22 \ensuremath{\mu\mathrm{s}}. The stretching
exponent falls to n \ensuremath{\approx} 0.58 (Fig. \ref{fig:6} f,g). This indicates a broader
distribution of characteristic times.The weak dependence on 976-nm power associates the change mainly with liquid contact. Simultaneous green and infrared illumination again shortens \ensuremath{T_1} (Fig. \ref{fig:5} h) and
reproduces the dry optical control.
Taken in sequence, Fig.~\ref{fig:5} \color{black} separates the trap's mechanical function from the hydrated interface's chemical function. This control permits the dose-dependent changes in Fig.~\ref{fig:6} to be interpreted at the material-liquid junction rather than assigned automatically to the trapping beam.Surface-mediated NV biochemical relaxometry provides precedent for indirect transduction \cite{ref26,ref27,ref28}.Independent orientation tracking defines a route to absolute angular
readout \cite{ref29}.
With the optical contribution constrained by Fig. \ref{fig:5} we treat the
accessible junction as a quasi-equilibrium charge-regulated
grand-canonical interface evaluated at nominal room temperature (298 K)
\cite{ref30}. Equation \ref{eq:main-3} connects aqueous bulk activity to a sensing plane
at potential \ensuremath{\psi_0}. Proton activity is dimensionless on the selected
standard state.At 298 K 59.2 mV corresponds to one local-pH unit under
this mean-field relation. The sign and magnitude of \ensuremath{\psi_0} remain free
because vaterite, PSS, nanodiamond coverage, ionic strength and
adsorbates all participate in charge regulation. Equilibration time and
temperature were not independently measured. Reversibility was also not
measured. Strict equilibrium is therefore a modelling assumption.
\begin{equation}
\begin{aligned}
\mu_{\mathrm{H,loc}}&=\mu_{\mathrm H}^{\circ}
+k_{\mathrm B}T\ln a_{\mathrm{H,bulk}}-e\psi_0,\\
\mathrm{pH}_{\mathrm{loc}}&=\mathrm{pH}_{\mathrm{bulk}}
+\frac{e\psi_0}{k_{\mathrm B}T\ln 10}.
\end{aligned}\label{eq:main-3}
\end{equation}
\\We coarse-grain proton-active vaterite and nanodiamond states into classes with areal density $\rho_s$ and occupation $f_s$.Strong polyelectrolyte PSS sulfonates are treated primarily as fixed charges and a counterion reservoir.Equation~\ref{eq:main-4} describes the interfacial grand potential per unit area. The bracketed logarithmic term is the dimensionless mixing free-energy function. Its physical configurational entropy has the opposite sign. The electrostatic density \ensuremath{u_{\mathrm{el}}} has units J m\ensuremath{^{-2}}. Here \ensuremath{\sigma} is the occupation-dependent charge density and \ensuremath{\partial}\ensuremath{u_{\mathrm{el}}}/\ensuremath{\partial}\ensuremath{\sigma} = \ensuremath{\psi_0}. Occupation and surface potential must therefore be solved self-consistently. This coupling is the charge-regulation mechanism rather than an independent fit of two variables.
\begin{equation}
\begin{aligned}
\frac{\Omega}{A}
&=\sum_s\rho_s\Bigg\{
f_s\left(\Delta G_s^{\circ}-\mu_{\mathrm{H,bulk}}\right)
\\
&\quad
+k_{\mathrm B}T
\left[
f_s\ln f_s+(1-f_s)\ln(1-f_s)
\right]
\Bigg\}
\\
&\quad
+u_{\mathrm{el}}(\sigma;I,\varepsilon),
\\[2mm]
\sigma
&=\sigma_{\mathrm{fix}}+e\sum_s\rho_s f_s,
\frac{\partial u_{\mathrm{el}}}{\partial\sigma}=\psi_0 .
\end{aligned}
\label{eq:main-4}
\end{equation}
\\Minimising equation (\ref{eq:main-4}) gives equation (\ref{eq:main-5}). The binomial variance \ensuremath{N_s}\ensuremath{f_s}(1
\ensuremath{-} \ensuremath{f_s}) is the conditional independent-site limit at fixed \ensuremath{\psi_0}.
Electrostatic feedback instead gives the full charge-regulated
covariance through the inverse Hessian of \ensuremath{\Omega}. This Hessian measures
interfacial susceptibility and can introduce cross-correlations plus
d\ensuremath{\psi_0}/d\ensuremath{\mu}H. The effective \ensuremath{\mathrm{p}K_{a,s}} values are distributed free-energy
coordinates of the composite interface. They are not the aqueous
bicarbonate--carbonate \ensuremath{\mathrm{p}K_{a2}} or a unique crystallographic group.\\ In DMEM
and ethanol H\ensuremath{_{n}}\ensuremath{_{o}}\ensuremath{_{m}} remains a mass-balance coordinate rather than local
activity. Buffering, adsorption, solvent activity and dose-dependent \ensuremath{\psi_0} therefore enter through an unresolved medium-specific mapping.
\begin{equation}
\begin{aligned}
f_s(\psi_0)
&=\left[
1+\exp\!\left\{
\beta\left(\Delta G_s^{\circ}
-\mu_{\mathrm{H,bulk}}
+e\psi_0\right)
\right\}
\right]^{-1}
\\
&=\left[
1+10^{\mathrm{pH}_{\mathrm{loc}}
-\mathrm{p}K_{a,s}^{\mathrm{eff}}}
\right]^{-1},
\\[2mm]
\operatorname{Var}(N_{\mathrm H,s}\mid\psi_0)
&=N_s f_s(1-f_s),
\\
\operatorname{Cov}(\mathbf N)
&=k_{\mathrm B}T
\left[
\nabla_{\mathbf N}^{2}\Omega
\right]^{-1}.
\end{aligned}
\label{eq:main-5}
\end{equation}

Thermodynamic availability affects \ensuremath{T_1} only when a relaxation-active bath
carries spectral weight at an allowed quantum transition. Equation\ref{eq:main-6} separates an occupation-gated fast interfacial term with correlation
time \ensuremath{\tau_e} from a chemical occupation-switching term with \ensuremath{\tau_{\mathrm{chem}}}. \ensuremath{A_{s,m}} and
\ensuremath{B_{s,m}} contain the number, geometry and squared transverse coupling of the
effective fluctuators and have units s\ensuremath{^{-}}\textsuperscript{2},s\ensuremath{^{-}}\textsuperscript{1}.L has units of seconds. The
switching term contributes directly at the NV transition only if
occupation changes a transverse relaxation-active field on that
timescale.The factor f(1 \ensuremath{-} f) is a thermodynamic susceptibility. It
does not prove that proton transfer occurs at the NV frequency. \ensuremath{\Gamma_{1,\mathrm{eff}}}
is a comparison metric for stretched ensemble decays. Microscopic rate
additivity is exact only for exponential Markovian relaxation.

\begin{equation}
\begin{aligned}
\Delta\Gamma_{1,\mathrm{eff},m}(u)
&=\sum_s\Bigg[
A_{s,m}f_s(u)
L(\omega_0,\tau_{\mathrm e,s,m})
\\
&\quad
+B_{s,m}f_s(u)\big(1-f_s(u)\big)
L(\omega_0,\tau_{\mathrm{chem},s,m})
\Bigg]
\\
&=\sum_s A_{s,m}f_s(u)
L(\omega_0,\tau_{\mathrm e,s,m})
\\
&\quad
+\sum_s B_{s,m}f_s(u)\big(1-f_s(u)\big)
L(\omega_0,\tau_{\mathrm{chem},s,m}),
\\[2mm]
L(\omega,\tau)
&=\frac{2\tau}{1+\omega^2\tau^2}.
\end{aligned}
\label{eq:main-6}
\end{equation}
For the single-Lorentzian closure the fast term is maximally weighted at
\ensuremath{\omega_0}\ensuremath{\tau_e} = 1. This gives \ensuremath{\tau_e} \ensuremath{\approx} 55 ps at the 2.87-GHz NV transition. This is a constraint on a downstream electronic, paramagnetic or defect-state fluctuator.It is neither a measured proton-exchange time nor evidence of direct proton nuclear-spin noise. Slow protonation may instead regulate the population, coupling or correlation time of that faster bath. Proton Zeeman precession at 0.5--0.8 mT is only 21--34 kHz. It lies far below the single-quantum resonance.Fig.\ref{fig:6} \color{black} tests this thermodynamic-kinetic transduction across three media.\\
\\Proton-responsive relaxometry was examined in three media (Fig. \ref{fig:6}).In the NaOH/HCl series, the initial pH 10.22 corresponds to {[}H\ensuremath{^{+}}{]} \ensuremath{\approx} 6.0 \ensuremath{\times} 10\ensuremath{^{-}}\textsuperscript{1}\textsuperscript{1} M(0.060 nM). Controlled additions of a 5.2-mM HCl stock solution produce monotonic \ensuremath{T_1} shortening. Lowering measured pH from 10.22 to 8.77 raises the reservoir proton chemical potential by kBT ln(10) \ensuremath{\times} 1.45 =
3.34 kBT. Regional fits over pH 10.22--9.86 and 9.79--8.77 give source-reported response metrics of 0.250 and 1.25 pH Hz\ensuremath{^{-1/2}}. The slope change is consistent with a heterogeneous carbonate-coupled window. It
does not establish a phase transition or microscopic \ensuremath{\mathrm{p}K_a}. The metrics require an acquisition-time noise model before they can be validated as sensitivities.
\\In open-air DMEM microlitre additions of pH 4.01 potassium hydrogen phthalate buffer change measured bulk pH only from 8.100 to 8.088 (Fig. \ref{fig:6} b). The independent variable is therefore nominal proton-equivalent
dose H\ensuremath{_{n}}\ensuremath{_{o}}\ensuremath{_{m}} rather than free interfacial activity. \ensuremath{T_1} decreases from 23.4 \ensuremath{\pm} 2.3 to 9.0 \ensuremath{\pm} 1.2 \ensuremath{\mu\mathrm{s}} at H\ensuremath{_{n}}\ensuremath{_{o}}\ensuremath{_{m}} = 10.7 \ensuremath{\mu\mathrm{M}}. This is equivalent to \ensuremath{\Gamma_{1,\mathrm{eff}}} increasing from 42.7 \ensuremath{\pm} 4.2 to 111.1 \ensuremath{\pm} 14.8 ms\ensuremath{^{-1}}. The change is \ensuremath{\Delta\Gamma_{1,\mathrm{eff}}} = 68.4 \ensuremath{\pm} 15.4 ms\ensuremath{^{-1}} and 2.60 \ensuremath{\pm} 0.43-fold. Yet the 0.012-pH-unit bulk change is only 0.0276 kBT (0.71 meV at 298 K). At fixed \ensuremath{\psi_0} and activity coefficients it changes any ideal binary-site occupation by at most 0.69 percentage points. The response therefore rules out a fixed-potential independent-site explanation based solely on the measured bulk-pH shift. It requires amplification through local composition and electrostatic feedback. Cooperative or structural changes, adsorption and bath kinetics may also contribute. The reported 6.46 \ensuremath{\pm} 0.04 \ensuremath{\mu\mathrm{M}} Hz\ensuremath{^{-1/2}} and 6.54 \ensuremath{\pm} 0.02 mpH Hz\ensuremath{^{-1/2}} values remain fit-derived response metrics rather than absolute local-pH calibrations.
\\In ethanol nominal additions of 2,817 and 5,333 \ensuremath{\mu\mathrm{M}} change \ensuremath{T_1} from 32.6 \ensuremath{\pm}3.2 \ensuremath{\mu\mathrm{s}} to 30.1 \ensuremath{\pm} 3.2 and 20.2 \ensuremath{\pm} 3.9 \ensuremath{\mu\mathrm{s}} respectively (Fig. \ref{fig:6} c).The corresponding \ensuremath{\Gamma_{1,\mathrm{eff}}} values are 30.7 \ensuremath{\pm} 3.0, 33.2 \ensuremath{\pm} 3.5 and 49.5 \ensuremath{\pm} 9.6 ms\ensuremath{^{-1}}. The first increment is unresolved.The endpoint trend is \ensuremath{\Delta\Gamma_{1,\mathrm{eff}}} = 18.8 \ensuremath{\pm} 10.0 ms\ensuremath{^{-1}} and is approximately 1.9 fit-standard-errors. It occurs at a nominal dose about 500-fold larger than the DMEM endpoint.The source-reported nominal-dose metric is (6.33 \ensuremath{\times} 10\textsuperscript{3} \ensuremath{\pm} 38) \ensuremath{\mu\mathrm{M}} Hz\ensuremath{^{-1/2}}. Acidity scales and activity coefficients are solvent dependent.Aqueous pH, \ensuremath{\mathrm{p}K_a} and equation \ref{eq:main-3} are therefore not transferred quantitatively to ethanol. Its electrode coordinate remains descriptive.
\\The ethanol control therefore shows that the response is not set by acid dose alone. The DMEM--ethanol contrast constrains the medium-dependent amplitudes and timescales in equation \ref{eq:main-6}. It is not an affinity,
selectivity, or free-energy ratio because nominal dose is not interfacial activity. Hydration and buffer capacity change together with ion dissociation, adsorption and screening.Surface potential and nanodiamond defect dynamics also change. Furthermore metastable vaterite may dissolve or restructure during acid exposure.Without reversibility,Ca\ensuremath{^{2+}} release or post-titration morphology that pathway cannot be separated from charge regulation.\\
The carbonate coordinate in Fig. \ref{fig:6} d supplies an aqueous reference rather than a fitted surface constant.Across the NaOH/HCl series {[}HCO\ensuremath{_3^{-}}{]}/{[}CO\ensuremath{_3^{2-}}{]} increases from 1.29 at pH 10.22 to 36.3 at pH 8.77 using \ensuremath{\mathrm{p}K_{a2}} = 10.33. DMEM occupies a narrow bicarbonate-dominated window.Ethanol is retained only on an apparent-pH coordinate. This alkaline coordinate is distinct from the near-neutral CO\ensuremath{_{2}}/HCO\ensuremath{_3^{-}} buffering axis used throughout many biological fluids. The solid
vaterite interface links the sensor to this pervasive inorganic-carbon chemistry but does not itself constitute a soluble buffer.Equations \ref{eq:main-3}-\ref{eq:main-6} therefore organise bulk chemical potential, charge regulation,
occupation statistics and resonance-filtered dynamics.The dashed curves in Fig. \ref{fig:6} remain empirical regional or Hill-like projections rather than global fits of the thermodynamic parameters.
These measurements establish a dose-dependent response in the measured particle series. They do not establish selectivity for H\ensuremath{^{+}} over changes in phthalate, potassium, ionic strength or carrier dissolution.An absolute pH sensor requires matched-buffer controls and acid--base reversibility. It also requires controlled equilibration time and temperature with defined CO\ensuremath{_{2}} conditions; measurement of surface or zeta potential, an acquisition-time noise budget; particle-level replication. \\
Within those limits the causal chain remains clear.Vaterite supplies the optomechanical frame and charge-regulated interface. The surrounding medium controls occupation and fluctuation kinetics.The surface-bound NV layer converts the resulting resonance-filtered bath into \ensuremath{\Gamma_{1,\mathrm{eff}}}.The carrier thus separates optical torque, anisotropic emission and resonance-filtered interfacial noise into modular transfer functions.The quantum transducer therefore remains physically connected to the liquid while the carrier organises its optical and thermodynamic boundary conditions.
\section{Conclusions}
We developed a vaterite-nanodiamond microcarrier that couples an
optically addressable body frame with orientation-dependent emission
redistribution and a charge-regulated chemical interface. Vaterite
thereby unites birefringent optical control with a carbonate-rich
chemical boundary. Under 976-nm trapping the hybrids retain ODMR and \ensuremath{T_1}
readout. They resolve 0--0.8 mT fields with a source-reported 78--144 \ensuremath{\mu\mathrm{T}}
Hz\ensuremath{^{-1/2}} response metric. In DMEM the 2.60 \ensuremath{\pm} 0.43-fold increase in \ensuremath{\Gamma_{1,\mathrm{eff}}}
is far larger than the maximum fixed-potential ideal-site occupation
change permitted by the measured 0.012-unit bulk-pH shift. The combined
data therefore support amplified medium-dependent interfacial
transduction rather than direct proton-spin resonance or a bulk-pH-only
response.
\\Furthermore the same carrier connects optics, quantum resonance and
interfacial thermodynamics without a fabricated cavity or fixed diamond
substrate.This modular architecture provides a route toward a universal
carbonate-based nanoplatform.Universality across analytes and biofluids
will still require selective functionalisation and calibration.The
present data establish orientation-stabilised proton-responsive sensing
rather than a fully calibrated angular lock, microscopic site model or
absolute pH measurement. Direct angular tracking, matched chemical
controls, reversibility,local-potential measurements and particle-level
calibration are the requirements for quantitative vector magnetometry
and chemical mapping in heterogeneous biofluids.
\section*{References}
\bibliography{references}
\setcounter{figure}{0}
\renewcommand{\thefigure}{A\arabic{figure}}
\vspace{2em}
\newpage
\begin{center}
    {\LARGE\bfseries Methods\par}
    \vspace{1em}
\end{center}
\renewcommand{\thefigure}{S\arabic{figure}}
\setcounter{figure}{-1}

Experimental procedures for synthesis, structural and optical characterisation, optical trapping, microwave delivery and spin readout are described below. The Supporting Information follows the physical sequence of the experiment: formation of the carrier and its interface, optical delivery and collection, host-material luminescence, electromagnetic response, and then magnetic and relaxometric measurements. This organisation links each data set to the material component that produces or transduces it.

\subsection*{Synthesis of vaterite--nanodiamond hybrids}
\label{Synthesis of vaterite--nanodiamond hybrids}
Vaterite microspherulites were prepared by mixing 0.1 M aqueous Na\ensuremath{_{2}}CO\ensuremath{_{3}}
and 0.1 M aqueous CaCl\ensuremath{_{2}} at a 1:5 volume ratio to a total volume of 40 ml. After 30 s, the precipitate was collected by centrifugation at 5,000 r.p.m. for 30 s and washed three times: once with deionised water and twice with ethanol. The product was dried at room temperature. Reaction time and reagent ratio were selected to favour spherical vaterite particles.

The rapid precipitation step creates the metastable vaterite morphology used as the carrier. For the present work, the relevant material properties extend beyond CaCO\ensuremath{_{3}} composition to the micrometre-scale spherical shape, rough porous exterior and optical anisotropy.The sphere provides a body that can be trapped and torqued; the surface texture provides attachment area and liquid access; and the birefringence provides the orientation-dependent mechanical and optical response examined in Appendix B \cite{ref15,ref16,ref18}.

To prepare the hybrids, 0.5 ml of a 1 mg ml\ensuremath{^{-}}\textsuperscript{1} suspension of 40-nm fluorescent nanodiamonds (Adamas Nano) was added to 10 mg of dried vaterite and sonicated for 1 h. Poly(sodium 4-styrenesulfonate) (PSS) was included during assembly to modify the surface charge and improve dispersion and optical trapping.

PSS is therefore treated as a functional part of the hybrid interface rather than only as a processing additive. Its fixed sulfonate groups and associated counterions modify electrostatic interactions among the calcium-carbonate surface, the nanodiamonds and the solvent. This interlayer can influence dispersion, nanodiamond retention, hydration and ion adsorption, all of which are relevant when the same particle is later used for optical trapping and chemical relaxometry.
\subsection*{\textbf{Physical basis of the hierarchical carrier}}
The hybrid contains coupled length scales with distinct functions.The 3--5-\ensuremath{\mu\mathrm{m}} vaterite body is large enough for stable optical trapping and produces a well-defined mesoscale optical boundary. Its rough, porous surface provides access for solvents and ions. The 40-nm nanodiamonds remain much smaller than the carrier and act as the quantum-sensing layer, while the NV centres within each nanodiamond sample only their local near-field environment. This hierarchy decouples gross particle motion from nanoscale spin transduction without separating the sensor from the liquid.
Surface localisation is central to all three functions of the platform. Mechanically, the nanodiamonds move with the vaterite body instead of rotating independently. Optically, surface emitters interact most strongly with the curved carrier boundary and can produce orientation-dependent collection. Chemically, the same location places the nanodiamond surface and its near-surface NV centres close to hydrated carbonate, PSS, adsorbed ions and solvent. Nanodiamonds buried deeply in the carrier would be less accessible to the liquid and would not represent the geometry analysed in the electromagnetic model.
\subsection*{Structural and luminescence characterisation}
SEM and TEM were used to establish particle shape, porosity and the presence of nanoscale material at the carrier boundary. EDS was then used as a compositional cross-check: calcium identifies the CaCO\ensuremath{_{3}}-rich
host, whereas a carbon-enriched periphery is consistent with the combined contribution of nanodiamonds and PSS. The interpretation does not rely on carbon mapping alone because the polymer also contains
carbon.
\begin{figure*}[t]
\centering
\includegraphics[width=\linewidth,height=1\textheight,keepaspectratio]{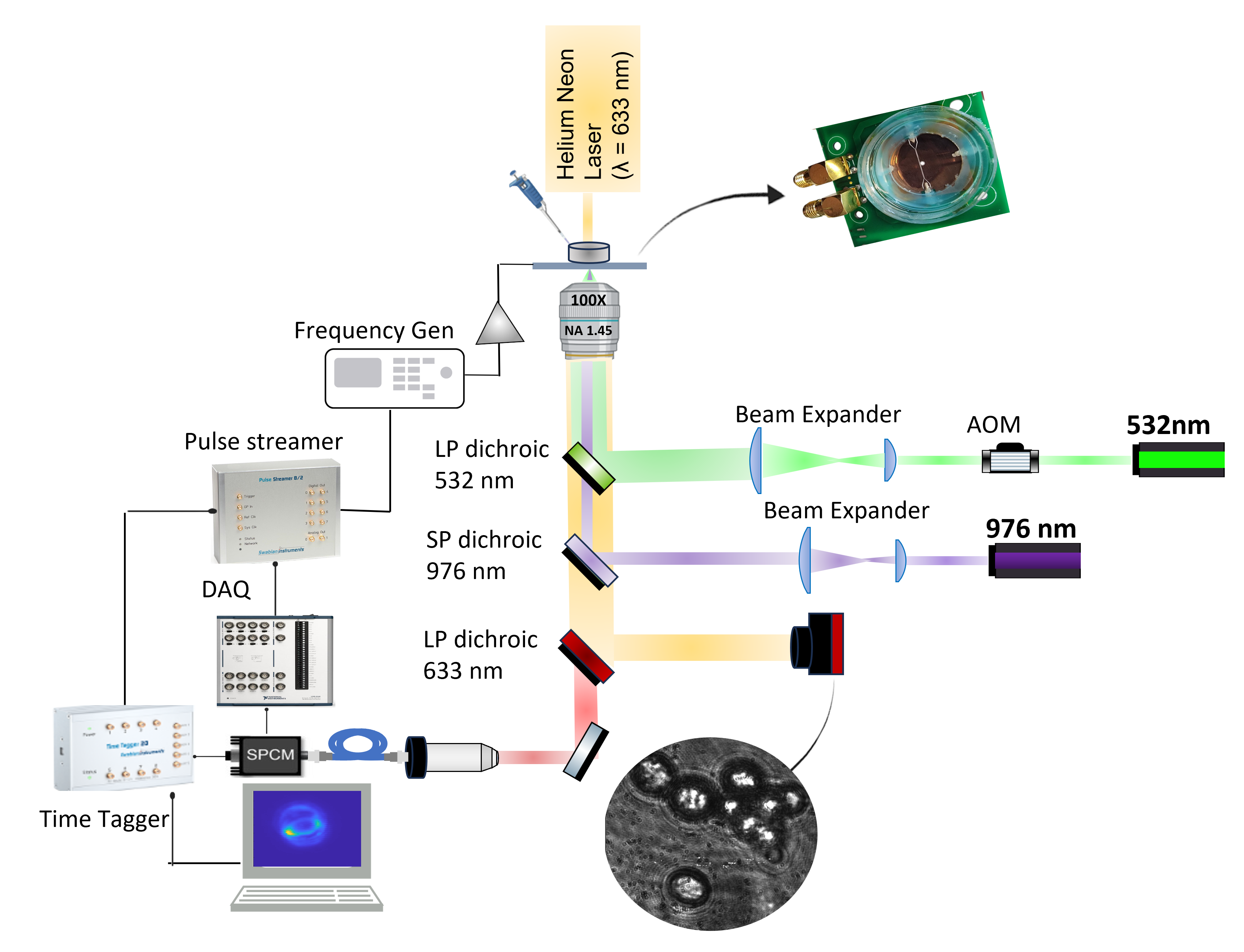}
\caption{ \emph{Schematic of the optical-trapping and confocal setup.}}
\label{fig:S0}
\end{figure*}
Confocal PL and CL provide the complementary optical assignment. Confocal PL reports the spatial distribution of red-emitting NV-bearing material under optical excitation. CL separates the intrinsic blue-violet host emission from the additional red surface band. Reading the electron-microscopy, elemental and luminescence data together is therefore more informative than treating any single map as proof of a continuous nanodiamond shell.
\subsection*{Optical trapping, confocal readout and microwave delivery}
The custom instrument combines a confocal microscope, optical tweezers and single-photon counting (Fig. S0). A 976-nm trapping beam and a 532-nm excitation beam were combined with dichroic mirrors and focused into the sample through a Nikon Plan Apochromat 100\ensuremath{\times}, NA 1.45 objective.
The 532-nm beam was used for NV spin initialisation and fluorescence readout. Bright-field images were obtained with a 633-nm He--Ne laser illuminating the sample from above. For the orientation-dependent measurements, a quarter-wave plate in the trapping path was rotated to vary the polarisation state; the resulting setting is reported as the polarisation-control angle.
Hybrid particles suspended in ethanol, water or D\ensuremath{_{2}}O, as specified for each experiment, were introduced onto a custom coplanar waveguide mounted on a coverslip with an open imaging window. A permanent magnet positioned above the sample supplied the static field used for ODMR measurements.
NV photoluminescence was collected through the same objective. A 976-nm short-pass filter rejected the trapping beam and a 650-nm long-pass filter selected the red NV emission. The signal was directed either to a Thorlabs camera for imaging or to an Excelitas single-photon counting module for time-resolved detection.
Microwaves were generated by an SRS SG396 source, amplified with a Mini-Circuits ZHL-5W-63S+ amplifier, and delivered through the coplanar waveguide. A Pulse Streamer 8/2 timing controller (8 digital channels and 2 arbitrary waveform generators) was used to synchronise the microwave, laser, and photon-counting sequences. Data acquisition and pulse control were implemented in MATLAB.
Fig. \ref{fig:S0} summarises the instrument in the order followed by the light and control signals. The 532-nm excitation and 976-nm trapping beams are expanded, combined by dichroic optics, and focused through the high-NA objective onto a hybrid located on the coplanar waveguide. Red photoluminescence returns through the same objective, passes the trapping-beam rejection and NV-emission selection filters, and is directed to either the camera or the single-photon counting module. The pulse generator synchronises the optical gates, microwave source, and photon-counting windows so that imaging, ODMR and \ensuremath{T_1} sequences use the
same spatially selected particle.

\section*{Data availability}
All data that support the findings of this study are available within the paper and its supplementary information. Any other relevant data are
available from the corresponding author upon request.
\samepage
\section*{Code availability}
The custom code developed for the numerical simulations and data analysis in this study, including the code used in conjunction with COMSOL Multiphysics, is available from the corresponding author upon reasonable request.
\section*{Acknowledgments}
The Ministry of Innovation, Science and Technology, Israel (Grant No 0008648 H.B. and N.B.).\\
\\
The Ministry of Innovation, Science and Technology, Israel (Grant No 7407 M.A.).\\
\\
We acknowledge Tel Aviv University for providing access to COMSOL Multiphysics through its institutional license, and thank Tal Carmon for facilitating this access.
\section*{Competing interests}
The authors declare no competing interests.
\end{document}


\onecolumngrid
\begin{center}
{\LARGE\bfseries
Supplementary Information: Biocompatible Vaterite Carriers Enable Multimodal Quantum Sensing with Nanodiamonds.
\par}
\vspace{1em}

{\large
Tamara Amro$^{1,2}$,
Mohammad Attrash$^{2,3}$,
Ali Droby$^{4}$,
Andrei Ushkov$^{3}$,\
Rotem Malkinson$^{1}$,
Pavel Penshin$^{1}$,
Amir Hen$^{1}$,\
Vjaceslavs Bobrovs$^{5}$,
Pavel Ginzburg$^{3}$,
Hani Barhum$^{2,3,*}$,
Nir Bar-Gill$^{1,6,7,*}$
\par}

\vspace{0.8em}

{\small
$^{1}$Dept. of Applied Physics, Rachel and Selim School of Engineering, Hebrew University, Jerusalem 91904, Israel\\
$^{2}$Triangle Regional Research and Development Center, Kfar Qara’ 3007500, Israel\\
$^{3}$School of Electrical Engineering, Tel Aviv University, Ramat Aviv, Tel Aviv 69978, Israel\\
$^{4}$Physics Program, Graduate Center, The City University of New York, New York, NY 10016, US\\
$^{5}$Institute of Telecommunications, Riga Technical University, Riga LV-1048, Latvia\\
$^{6}$The Racah Institute of Physics, The Hebrew University of Jerusalem, Jerusalem 91904, Israel\\
$^{7}$The Center for Nanoscience and Nanotechnology, The Hebrew University of Jerusalem, Jerusalem 91904, Israel\\
\par}

\end{center}

\vspace{1cm}

\FloatBarrier

\FloatBarrier\section*{Supplementary Note A: Vaterite optical characterization}
\renewcommand{\thefigure}{A\arabic{figure}}
\setcounter{figure}{0}
\renewcommand{\theequation}{S\arabic{equation}}
 \setcounter{equation}{0}
Figure~\ref{fig:A1} is read from the spatial map to the spectrum. Panel a shows the CL distribution from vaterite under 30-keV electron-beam excitation, and panel b shows the corresponding emission spectrum with a prominent maximum near 405 nm. This blue-violet band provides the host-material reference used in the main-text comparison.Its spectral separation from the red NV band is important: the additional red emission observed at the hybrid surface in Fig.\ref{fig:fig2} is assigned through the combined spatial and spectral contrast, rather than through brightness alone With the intrinsic host emission identified, Appendix B turns to the second physical role of the same material: how a birefringent,micrometre-scale vaterite body changes the propagation and collection of light emitted by a surface-bound nanodiamond.
\begin{figure}[h]
\centering
\includegraphics[width=0.7\textwidth]{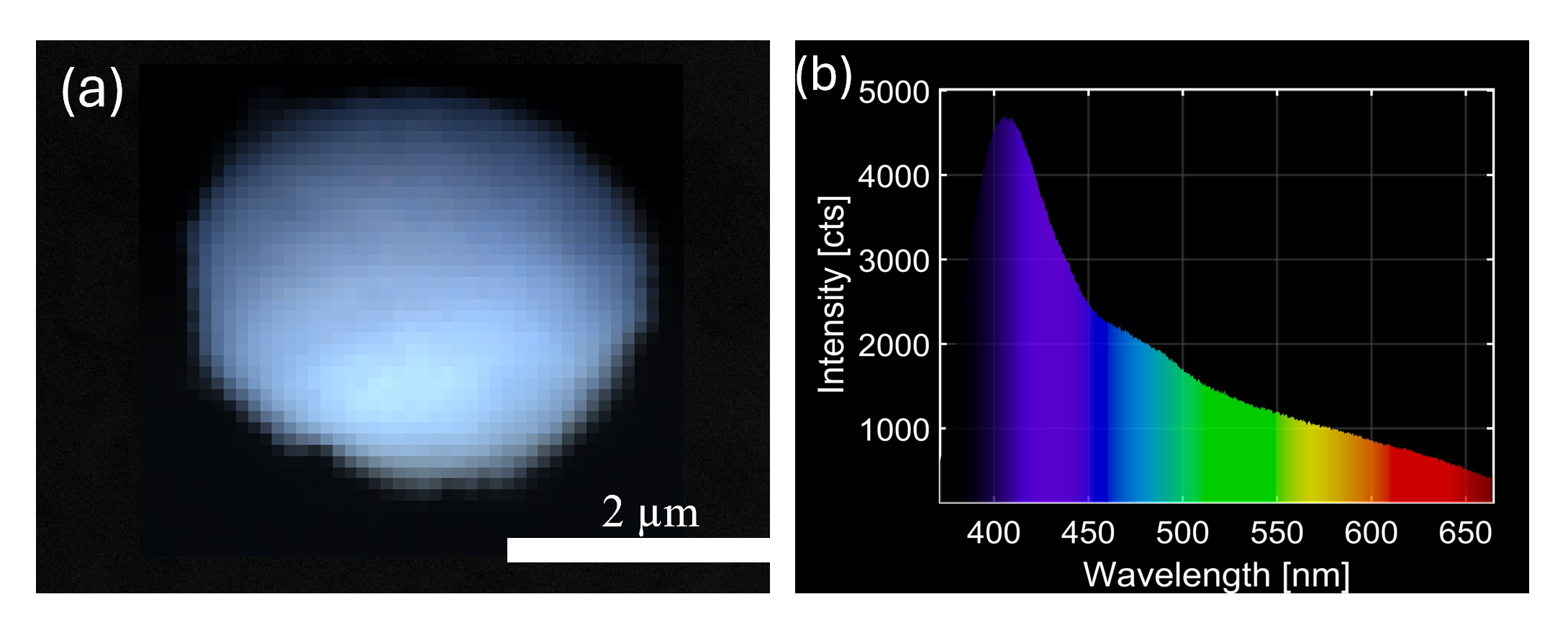}
\caption{Cathodoluminescence characterization of vaterite. (a) CL maps were acquired using a 30~keV electron beam at 17~nA. (b) CL spectrum of vaterite, showing a prominent peak at 405~nm corresponding to the intrinsic luminescence of the vaterite host.}
    \label{fig:A1}
\end{figure}
\label{Appendix A}

\FloatBarrier\section*{Supplementary Note B: Electromagnetic simulations and orientation-dependent photoluminescence}

\renewcommand{\thefigure}{B\arabic{figure}}
\setcounter{figure}{0}

\section*{B1.FDTD simulations}
\renewcommand{\thefigure}{B1.\arabic{figure}}
\setcounter{figure}{0}
FDTD simulations were used to determine how a birefringent vaterite microsphere redistributes emission from a surface-bound nanodiamond.The nanodiamond was represented as a 40-nm diamond sphere containing an electric dipole, and the surrounding liquid was represented as water. Near-field maps were evaluated at 680 nm unless otherwise noted, far-field patterns at 650 nm, and a calculation at 640 nm confirmed the same qualitative directionality. The vaterite body was represented by the anisotropic permittivity tensor described in \cite{ref18}., so the model retains the distinction between optical-axis and transverse responses that underlies both torque and directional emission.
The real particles are porous spherulites with surface roughness,nanodiamond clustering and a PSS-containing interface. The calculation intentionally replaces that complexity with an effective anisotropicn continuum sphere. It therefore tests the physical consequences of carrier size, emitter position and dipole orientation, but it does not reproduce pore-scale scattering, the coverslip, every nanodiamond--vaterite contact or the exact collection numerical aperture. The results are consequently interpreted as trends in angular redistribution and local-density-of-states modification rather than as absolute predictions of detected counts.\\
The simulations compare dipoles in the same spatial order used in the figures: at the sphere surface, at r = R/2 and at the centre. Within each position, x-, y- and z-oriented dipoles are evaluated. This arrangement separates two physical variables. Moving the dipole changes its overlap with the carrier boundary and internal modes, whereas rotating the dipole changes how its field projects onto the anisotropic material tensor. Surface dipoles aligned with the model optical axis show the strongest forward lobe; directionality decreases for emitters placed deeper in the sphere, and central emitters approach a more symmetric pattern.\\
The total-power ratio was defined as the power radiated by the same dipole in the vaterite-nanodiamond geometry divided by that in homogeneous water. The calculated surface-geometry values are approximately 0.30 and remain below unity. Accordingly, the principal result is angular redistribution of emission; the simulations do not show net radiative-rate enhancement relative to the chosen reference.\\
Three dipole positions were analysed: the microsphere surface, r = R/2 and the geometric centre. At each position, dipoles were oriented along x, y and z. Multiple surface positions were sampled where computational symmetry permitted.
The simulation volume contained the vaterite sphere and surrounding water. Symmetric or antisymmetric boundaries were used only when permitted by the dipole configuration; the remaining boundaries were perfectly matched layers. A 2-nm mesh was used around the nanodiamond and a 10-nm mesh in the outer domain.\\
Fields were sampled on three orthogonal planes through the dipole. For surface emitters, x- and y-oriented dipoles produce broad lateral distributions, whereas a z-oriented dipole produces a pronounced forward lobe (Fig.\ref{fig:B1.1} a-c). The latter is consistent with anisotropy-assisted focusing at the sphere boundary~\cite{ref16}.\\
At r = R/2, the directional structure is weaker (Fig.\ref{fig:B1.1} d-f). At the centre, all three orientations are more symmetric because the local geometry approaches spherical symmetry (Fig.\ref{fig:B1.1} g-i).

These calculations show that both dipole position and orientation control the emission pattern; they do not establish a uniform enhancement for an ensemble of randomly oriented surface NV centres.

\begin{figure}[h]
\centering
\includegraphics[width=1\textwidth]{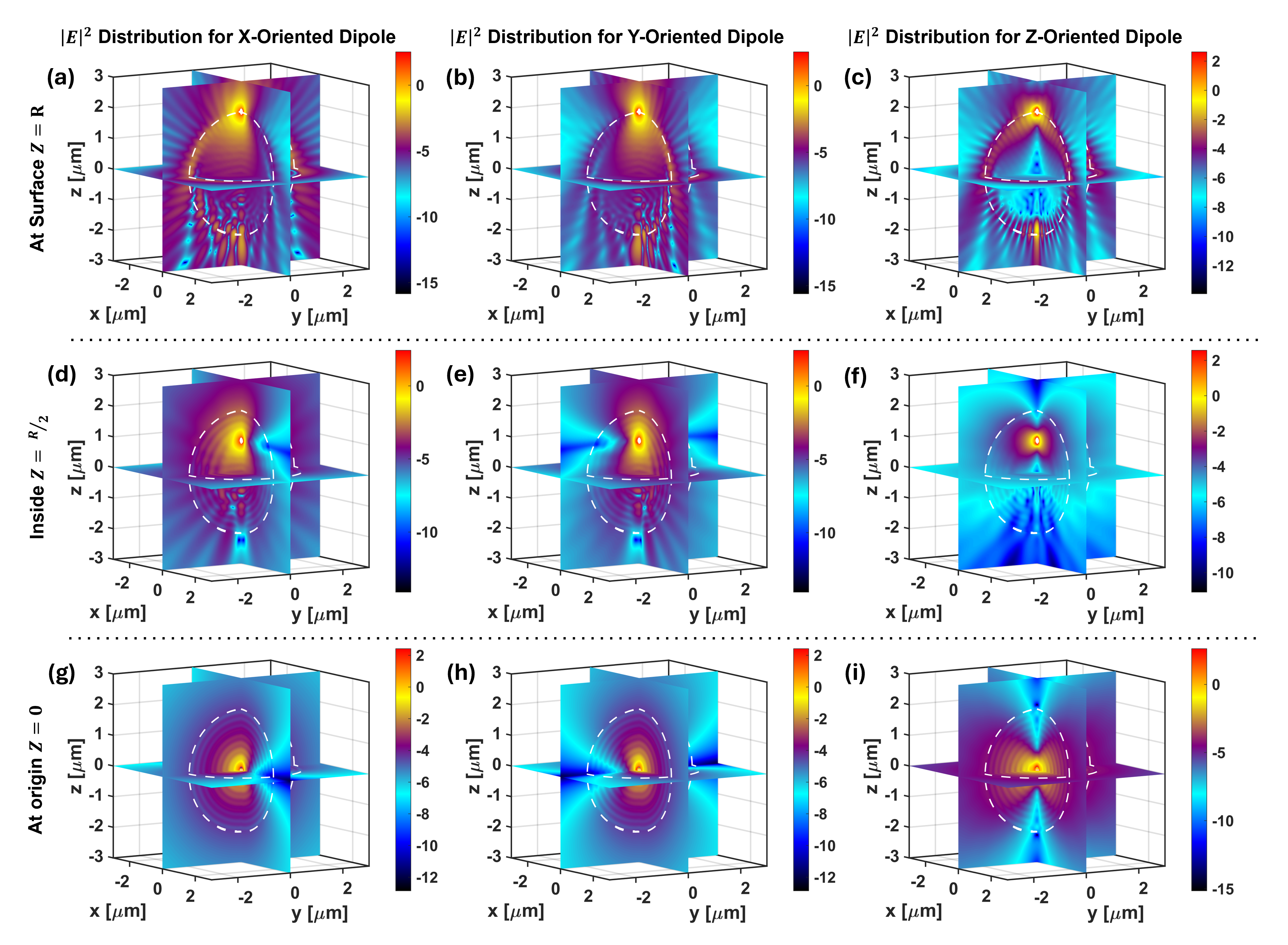}
\caption{\textbf{Electric field intensity ($|E|^2$) distributions for nanodiamond dipoles at different depths within a vaterite microsphere: }(a–c) surface, (d–f) interior (half of the radius), and (g–i) center. Dipoles are oriented along the x, y, and z axes, respectively.}
    \label{fig:B1.1}
\end{figure}
\begin{figure}[h]
\centering
\includegraphics[width=1\textwidth]{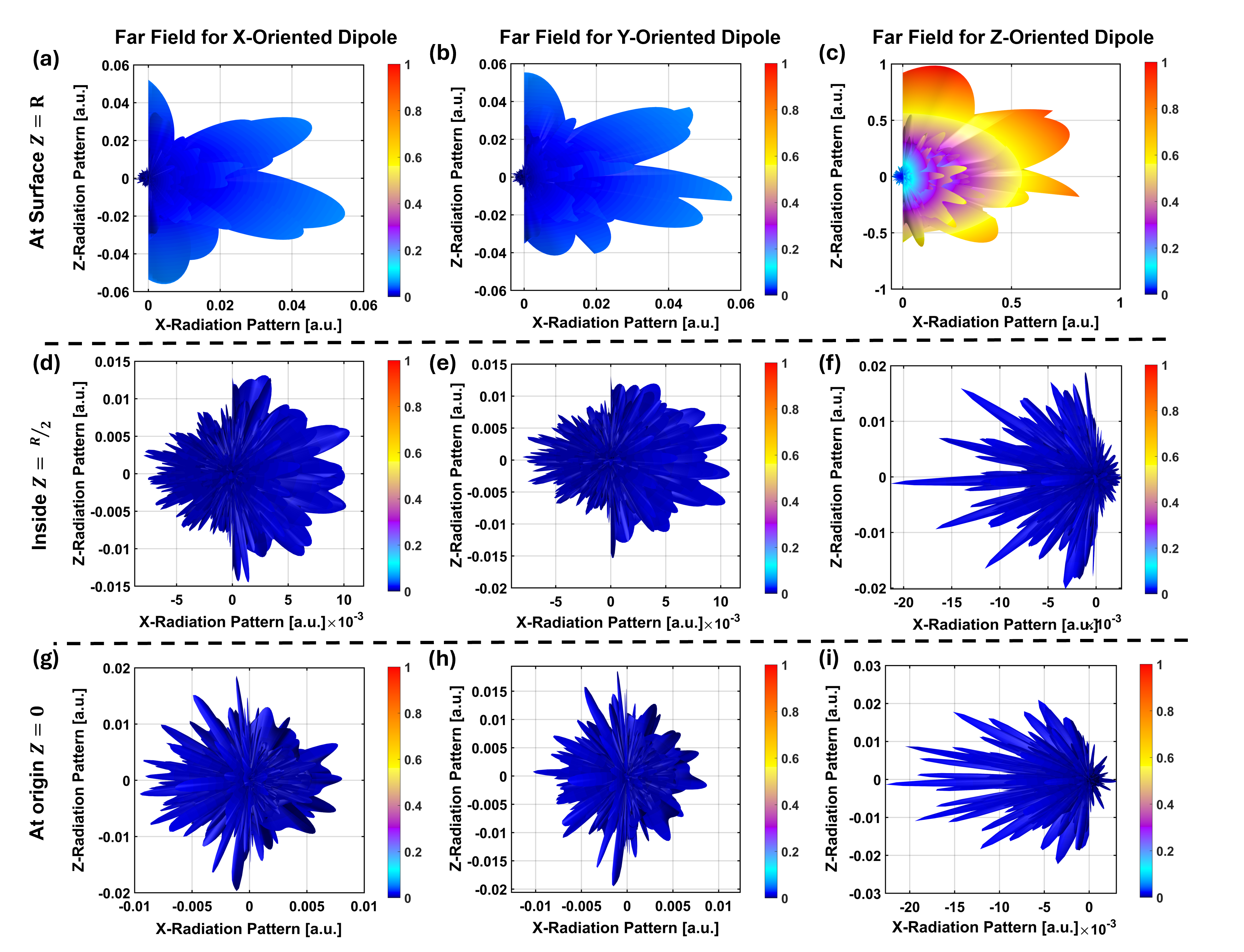}
\caption{\textbf{Far -Fireld simulation for nanodiamond dipoles at different depths within a vaterite microsphere: }(a–c) surface, (d–f) interior (half of radius), and (g–i) center. Dipoles are oriented along the x, y, and z axes, respectively.}
    \label{fig:B1.2}
\end{figure}
\begin{figure}[h]
\centering
\includegraphics[width=1\textwidth]{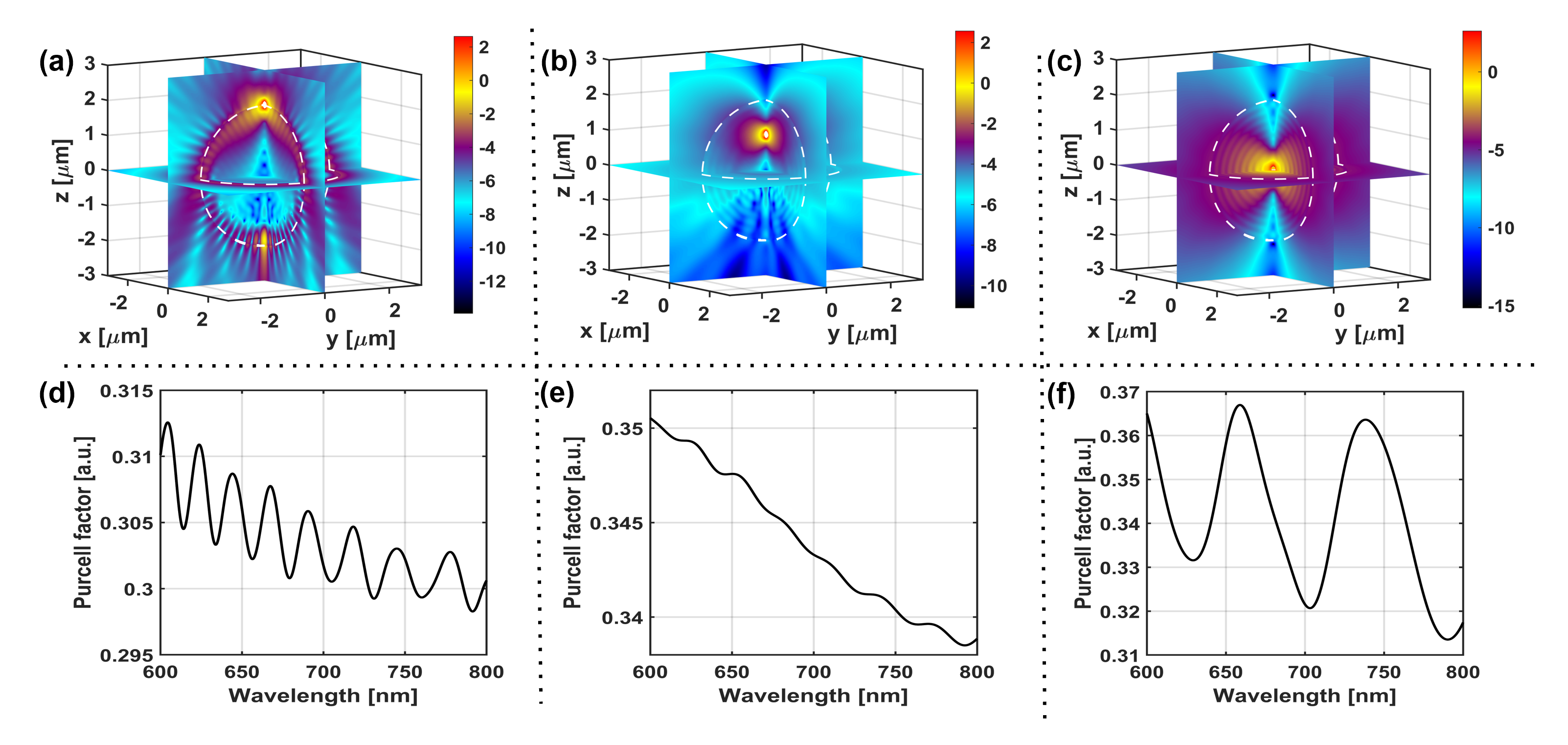}
\caption{(a–c) Electric field intensity ($|E|^2$) distributions for a z-oriented dipole located at (a) the microsphere surface, (b) the interior ($r = R/2$), and (c) the microsphere center. (d–f) Corresponding Purcell factor spectra averaged over all dipole orientations for dipoles placed at (d) the surface, (e) the interior ($r = R/2$), and (f) the center of the microsphere.}
    \label{fig:B1.3}
\end{figure}
\begin{figure}[h]
\centering
\includegraphics[width=0.8\textwidth]{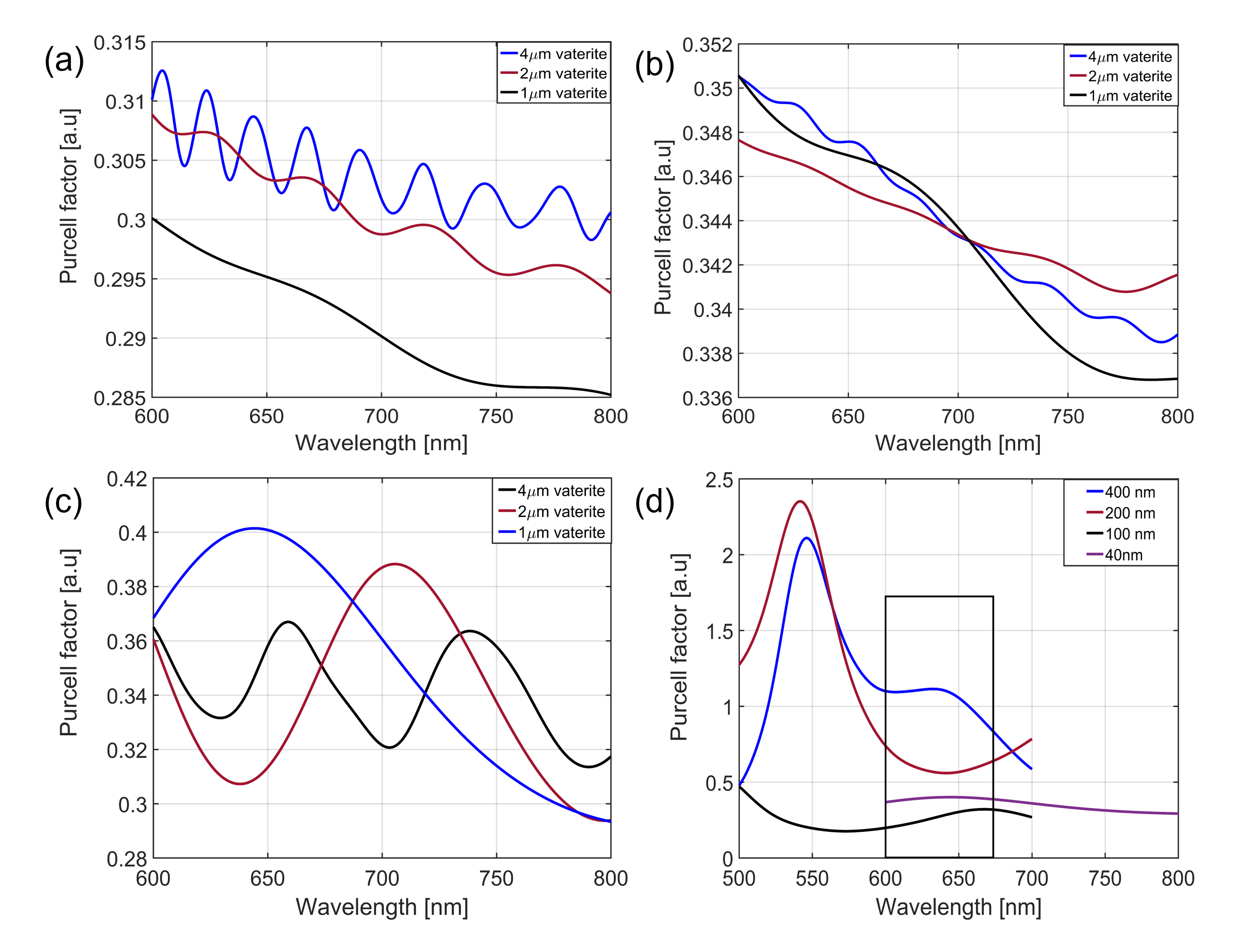}
\caption{\textbf{ Purcell factor simulations for 40\,nm nanodiamonds (NDs) embedded in vaterite particles of different diameters when the NDs are located at} (a) the particle surface ($r=R$), (b) inside the particle at $r=R/2$, and (c) the particle center ($r=0$). (d) Purcell factor spectra for NDs of different diameters positioned at the center of a 4\,\textmu m vaterite particle.}
    \label{fig:B1.4}
\end{figure}
\begin{figure}[h]
\centering
\includegraphics[width=0.8\textwidth]{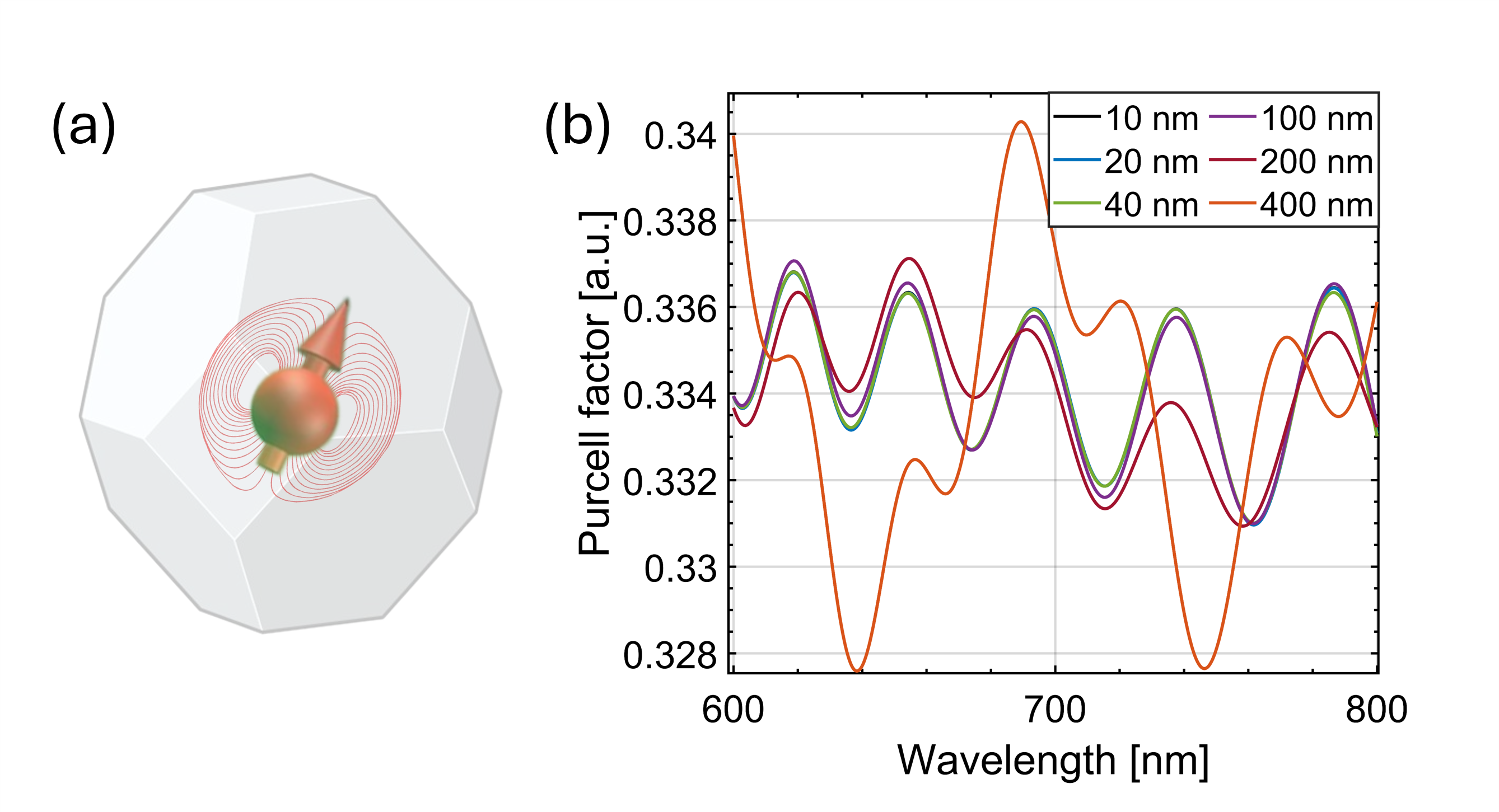}
\caption{\textbf{ Reference Purcell‑factor simulations.} (a) Schematic illustration of a nanodiamond containing a single electric dipole positioned at its geometric centre. (b) Corresponding Purcell‑factor spectra for several nanodiamond diameters, showing how emission‑rate enhancement varies with particle size.}
    \label{fig:B1.5}
\end{figure}

After the near-field distributions, near-to-far-field transforms were used to calculate the angular radiation patterns for the same positions and orientations. The rows of Fig.\ref{fig:B1.2} proceed from the surface (a--c), to r = R/2 (d--f), to the centre (g--i), and the columns proceed from x- to y- to z-orientated dipoles.\\
In the surface row, the z-orientated dipole produces the most pronounced forward lobe, whereas the x- and y-orientated dipoles are broader (Fig.\ref{fig:B1.2} a-c). At r = R/2, the angular structure remains orientation-dependent but is less concentrated (Fig.\ref{fig:B1.2} d-f). At the centre, all three patterns are comparatively symmetric
(Fig.\ref{fig:B1.2} g-i). Because each panel is normalised independently, the figure compares angular form rather than absolute collected power. The agreement between the near- and far-field sequences shows that the boundary-localised field structure is carried into the radiation zone, where it can affect the fraction of light entering the
objective.
\\
The near- and far-field calculations establish directionality, but directionality is not the same as an increased spontaneous-emission rate. The next calculation therefore evaluates the total radiated-power ratio relative to a clearly defined homogeneous-water reference.We also calculated the Purcell factor relative to an identical dipole in homogeneous water: \\
For an identical classical dipole, this is equivalent to the total-power ratio
\begin{equation}
F_{\mathrm P}=\frac{\Gamma}{\Gamma_{\mathrm{hom}}}.\label{eq:s1}
\end{equation}
The dipole moment and emission frequency are identical in the two calculations, so the total-power ratio describes modification of the local density of optical states relative to the stated reference.

\begin{equation}
F_{\mathrm P}=\frac{P_{\mathrm{tot}}}{P_{\mathrm{hom}}}.\label{eq:s2}
\end{equation}
Interior and central dipoles show stronger spectral modulation because they overlap more symmetrically with internal sphere modes. These configurations are parameter studies and do not represent the experimentally inferred surface-bound nanodiamond layer.\\
Fig. \ref{fig:B1.3} places the spatial and spectral comparisons side by side. Panels a--c show the z-oriented field distribution at the surface, at r = R/2 and at the centre, respectively. Panels d--f then show the corresponding orientation-averaged Purcell-factor spectra in the same positional order. The surface case is the experimentally
relevant geometry and remains below the homogeneous-water reference; the larger oscillations at interior and central positions illustrate stronger coupling to internal sphere modes but do not describe the measured surface-decorated particles.
\\
Fig.\ref{fig:B1.4} tests geometric parameters in panel order. Panels a--c vary the vaterite-sphere diameter for a 40-nm nanodiamond placed at the surface, at r = R/2 and at the centre, respectively. Increasing diameter changes the number and spectral positions of the mesoscale modes; the surface configurations remain weakly modulated, while interior and central configurations show larger oscillations. These curves compare model geometries and are not evidence that the experimental particle size was tuned during a single measurement.
Because every value is interpreted relative to the stated reference, a larger value within this set is described as stronger local-density-of-states modulation, not automatically as an enhancement above the reference. 
\\Panel d of Fig.\ref{fig:B1.4} varies the nanodiamond diameter for an emitter at the centre of a 4 \ensuremath{\mu\mathrm{m}} vaterite sphere. Changing the diamond size changes the local refractive-index distribution around the dipole and shifts its overlap with the microsphere modes. This centre-position parameter sweep is exploratory; the experimentally inferred nanodiamonds are located predominantly at the carrier surface.
These centre-position calculations are exploratory; the surface-localised configuration is the one relevant to the measured hybrids.
\\
As a reference baseline, we additionally simulated the Purcell factor for isolated NDs of varying diameters (10--400\,nm) embedded in a homogeneous water environment (see Fig.~\ref{fig:B1.5} (b), without the presence of the vaterite microsphere. The ND-alone simulations yield Purcell factors that are larger than those obtained for vaterite--ND hybrid particles. This is primarily due to the high refractive index of diamond ($n \approx 2.4$) relative to the surrounding medium, which strongly confines the electromagnetic field within and near the ND, thereby increasing the local density of optical states (LDOS) in the vicinity of the emitter.

For small NDs (10--100\,nm), the size is deeply subwavelength, and the Purcell spectra show nearly identical behavior across this range because the electromagnetic response is dominated by the refractive-index contrast rather than geometric resonances. Noticeable spectral modulation appears only for larger NDs ($\geq 200$\,nm), where the size parameter becomes comparable to the wavelength, allowing weak internal resonances to form.

In contrast, in vaterite--ND hybrid particles, the lower refractive index of the vaterite microsphere ($n \approx 1.6$) relative to diamond reduces the overall electromagnetic confinement, redistributing optical modes and resulting in a lower Purcell factor compared to bare NDs. This demonstrates that the refractive-index contrast between the emitter and its surrounding environment plays a dominant role in determining emission-rate enhancement, with the hybrid system showing more complex spectral features due to mode hybridization rather than simple additive enhancement.

Similar refractive-index-driven emission-rate enhancement in dielectric nanoparticles and nanodiamonds has been reported previously for NV centers and other solid-state emitters.

\section*{B2. Experimental orientation-dependent photoluminescenc}
The optical consequence of the carrier was then tested experimentally. Orientation-dependent PL was recorded while the quarter-wave plate in the trapping path was rotated.Fig \ref{fig:B2} is organised by particle from left to right. For each particle, the upper row shows a continuous scan, which captures the dynamic periodic response but is affected by non-uniform rotation speed; the lower row shows discrete polarisation-control settings, with 20-s averaging at each point over a complete 360\ensuremath{^{\circ}} cycle.

All three particles show a periodic response with two maxima separated by approximately 180\ensuremath{^{\circ}} and peak-to-trough changes of about 10--40\% ( Fig.\ref{fig:B2}). The repetition across particles connects the single-particle trace in Fig.\ref{fig:fig4} d to a broader material response, while the unequal maxima and different modulation depths reveal the effect of particle-shape and nanodiamond-coating heterogeneity. Because the physical particle angle was not tracked independently, these data demonstrate reproducible polarisation-dependent PL and an optically addressable body axis, not a calibrated torsional stiffness.
\renewcommand{\thefigure}{B2.\arabic{figure}}
\setcounter{figure}{0}
\begin{figure}[h]
\centering
\includegraphics[width=1\textwidth]{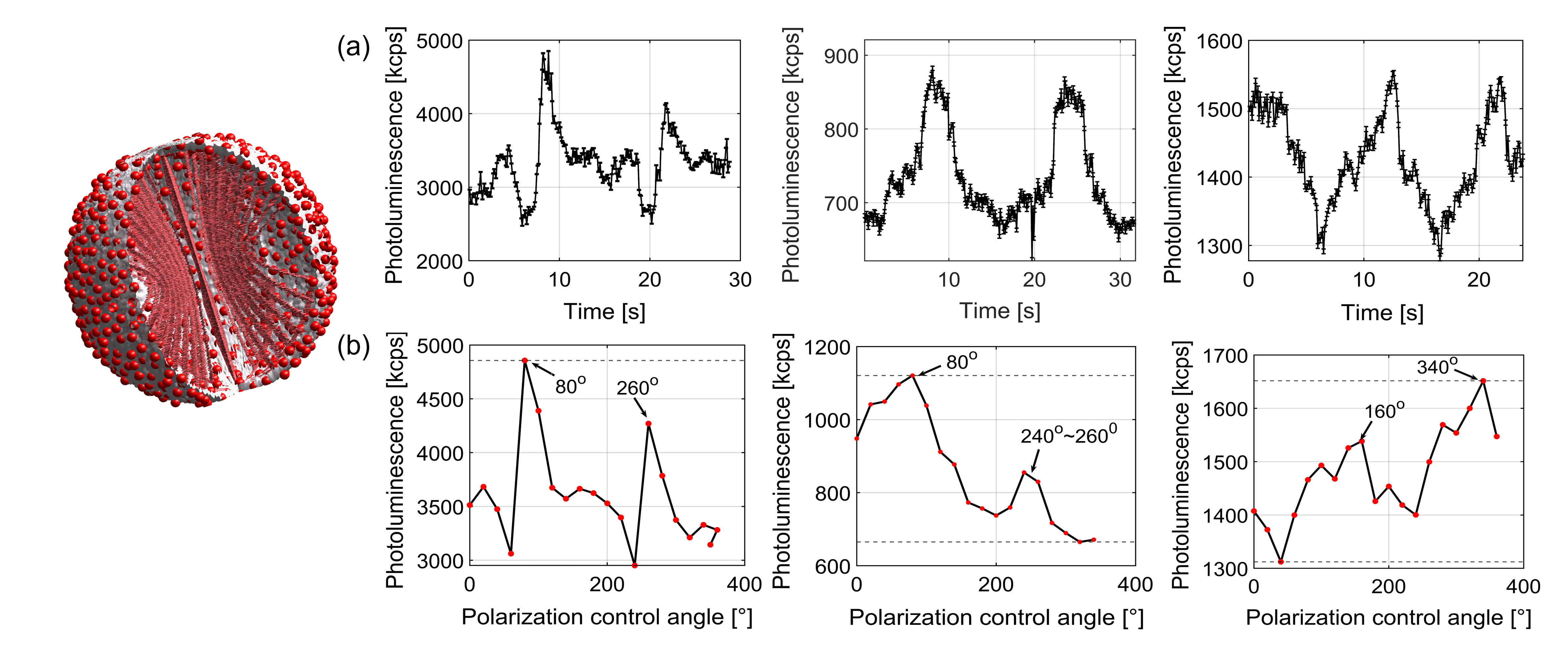}
\caption{\textbf{Polarisation-dependent PL from three trapped vaterite-nanodiamond hybrids.} (a) Continuous quarter-wave-plate rotation. (b) Discrete settings with 20-s averaging at each point.}
    \label{fig:B2}
\end{figure}

\textbf{Appendix B therefore separates two optical effects of the material: the carrier redistributes emission directionally, and the birefringent body produces a reproducible polarisation-dependent response in the trap. Appendix C next tests whether this optical control remains compatible with NV spin spectroscopy and interfacial relaxometry.}

\label{Appendix B}

\FloatBarrier\section*{Supplementary Note C: Quantum Sensing of Magnetic Fields and Local Chemical Environment through $T_1$ Relaxation}

The spin data are presented from material baseline to integrated device response. Supplementary Note C0 characterises the starting 40-nm nanodiamonds, Appendix C1 adds field-dependent spectra from trapped hybrids, and Supplementary Note C2 develops the effective description of the hydrated vaterite/PSS/nanodiamond interface. This progression distinguishes intrinsic nanodiamond heterogeneity from changes introduced by assembly, trapping and the surrounding medium.
\section*{C0. NV spin Hamiltonian, coupling fields and relaxometry}

\renewcommand{\thefigure}{C0.\arabic{figure}}
\setcounter{figure}{0}
A 40-nm nanodiamond contains a large surface-to-volume fraction, so strain, surface termination, paramagnetic defects and local charge fluctuations can vary substantially among particles. The measured PL, ODMR and \ensuremath{T_1} signals are therefore ensemble observables over multiple NV orientations and local environments within each selected fluorescent spot. This material heterogeneity motivates both the transverse-splitting term in ODMR fits and the stretched-exponential form used for relaxation.

NV\ensuremath{^{-}} centres have an optically addressable spin-triplet (S = 1) ground state. In frequency units, the minimal Hamiltonian is Supplementary Equation \ref{eq:s3} \cite{ref17}.

\begin{equation}
\frac{\widehat{\mathcal H}}{h}
=D\widehat S_z^2+E\left(\widehat S_x^2-\widehat S_y^2\right)
+\frac{\gamma_e}{2\pi}\mathbf B\cdot\widehat{\mathbf S}
+\frac{\widehat{\mathcal H}_{\mathrm{hf}}}{h}+\cdots .
\label{eq:s3}
\end{equation}

Here D and E are expressed in hertz, \ensuremath{\gamma}\ensuremath{_{e}} is in radians s\ensuremath{^{-1}} T\ensuremath{^{-1}}, and B is the local magnetic field. Hyperfine and additional strain/electric-field terms are collected in the hyperfine Hamiltonian and the omitted terms. ODMR uses the field-dependent transition frequencies, whereas \ensuremath{T_1} relaxometry measures transitions driven by fluctuating fields.

\begin{equation}
\Gamma_1^{\mathrm{micro}}
=\Gamma_{1,0}+\sum_{\alpha}g_{\alpha}^{2}
\left[S_{\alpha}(\omega_+)+S_{\alpha}(\omega_-)\right].\label{eq:s4}
\end{equation}
Here \ensuremath{\alpha} indexes effective magnetic, electric, strain or defect-state
noise channels; each channel has its corresponding spectral density
under the convention used in the fit, and \ensuremath{\omega_+} and \ensuremath{\omega_-} are the two allowed
transition frequencies. This form avoids assigning the chemical response
to a magnetic channel before the microscopic fluctuator is identified.

\begin{equation}
f_{\pm}\simeq D\pm\frac{\gamma_e}{2\pi}B_{\parallel}.\label{eq:s5}
\end{equation}

For approximately axial fields small compared with D, the transition
frequencies follow Supplementary Equation (S5), with \ensuremath{\gamma}\ensuremath{_{e}}/2\ensuremath{\pi} as the
electron gyromagnetic ratio in frequency units and \ensuremath{B_{\parallel}} as the axial field. Transverse fields and strain were treated through the fitted resonance positions rather than this approximation. In liquids, Brownian motion changes the NV-axis projection and optical collection \cite{ref29}; the hybrid carrier reduces these variations by providing a stable body frame.

Longitudinal relaxation was fitted with the stretched-exponential model described below. Near-surface paramagnetic defects and charge fluctuations can contribute to \ensuremath{T_1} in nanodiamonds \cite{ref5}.In the present hybrid, proton activity is treated as a regulator of the interfacial noise spectrum, not as direct nuclear-spin noise.

\begin{figure}[!htbp]
\centering
\includegraphics[width=1\linewidth,height=0.80\textheight,keepaspectratio]{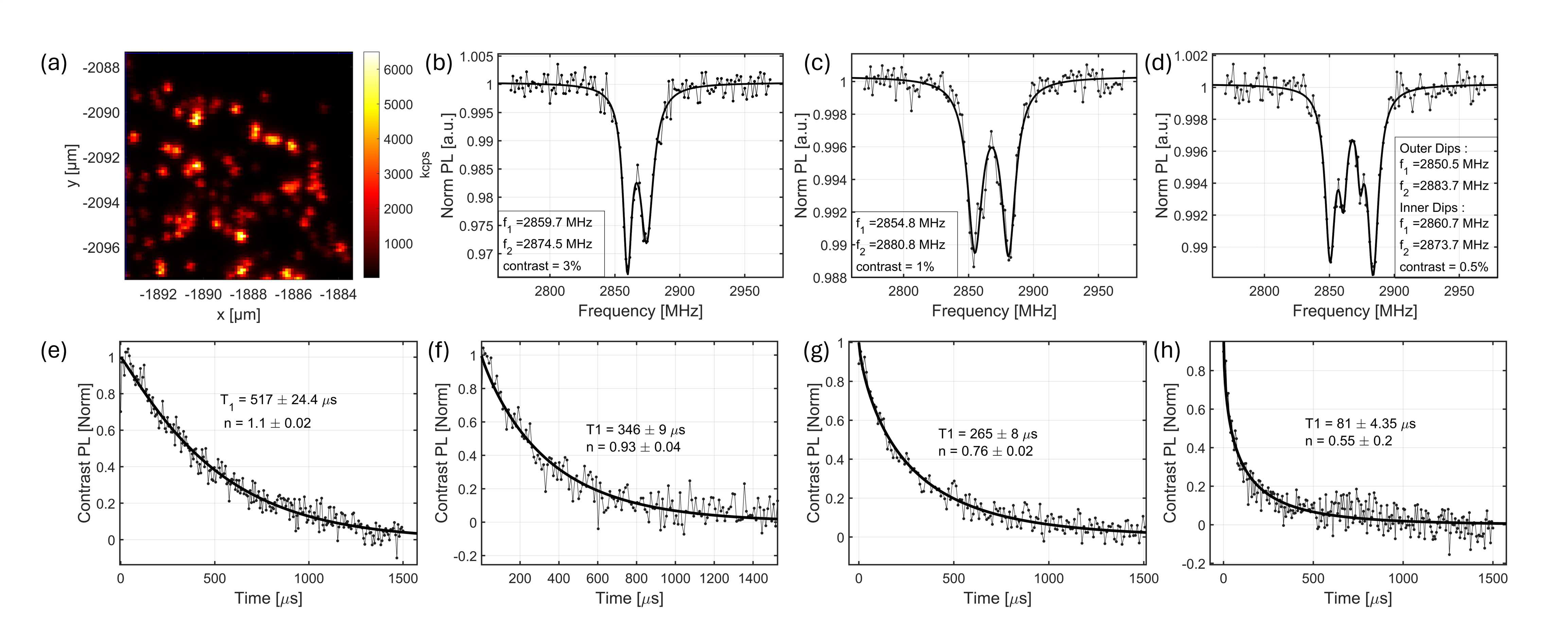}
\caption{\textbf{Spin characterisation of 40-nm nanodiamonds.} (a) Confocal PL map. (b--d) Continuous-wave ODMR spectra at increasing applied field. (e--h) Longitudinal-relaxation measurements from four nanodiamond spots.}
\label{fig:C0.1}
\end{figure}

Fig.\ref{fig:C0.1} presents the starting-nanodiamond baseline in measurement order. Panel a is the confocal PL map used to select fluorescent spots. Panels b--d are continuous-wave ODMR spectra at increasing applied field and show how resonance position, transverse splitting and the contribution of different NV orientations vary among the selected material. Panels e--h are \ensuremath{T_1} decays from four nanodiamond spots. Together, the panels show that the as-received 40-nm material has a broad distribution of optical and spin properties before it is attached to vaterite.
\\
The measured \ensuremath{T_1} values span approximately 60--550 \ensuremath{\mu\mathrm{s}}. Each decay was
fitted with

\begin{equation}
C(\tau)=C_0+A\exp\!\left[-\left(\frac{\tau}{T_1}\right)^n\right].\label{eq:s6}
\end{equation}

Here C(\ensuremath{\tau}) is the normalised PL contrast, \ensuremath{T_1} is the characteristic relaxation time and n is the stretching exponent. Values of n near 1 indicate a narrower distribution of local rates; smaller n indicates greater heterogeneity. Because \ensuremath{T_1} and n can be correlated in noisy ensemble fits, the quoted uncertainties are fit and should not be interpreted as particle-to-particle reproducibility.

\subsection*{C1. Magnetic-field sensing of trapped hybrids}
\renewcommand{\thefigure}{C1.\arabic{figure}}
\setcounter{figure}{0}
Supplementary Fig. \ref{fig:C1} follows the applied field sequentially from panel a to panel d: 0, 0.58, 0.63 and 0.77 mT (0, 5.8, 6.3 and 7.7 G). The zero-field panel establishes the trapped-particle line shape,and the subsequent panels show increasing separation and overlap of the broad ensemble resonances. Contrast decreases as the field increases because several NV orientations and strain-broadened features contribute simultaneously; residual angular motion within the trap may add further averaging. The data demonstrate detectable field-dependent splitting over the measured range without assigning biological relevance to the field magnitude in the absence of a specified source and distance.
\begin{figure}[!htbp]
\centering
\includegraphics[width=\linewidth,height=0.60\textheight,keepaspectratio]{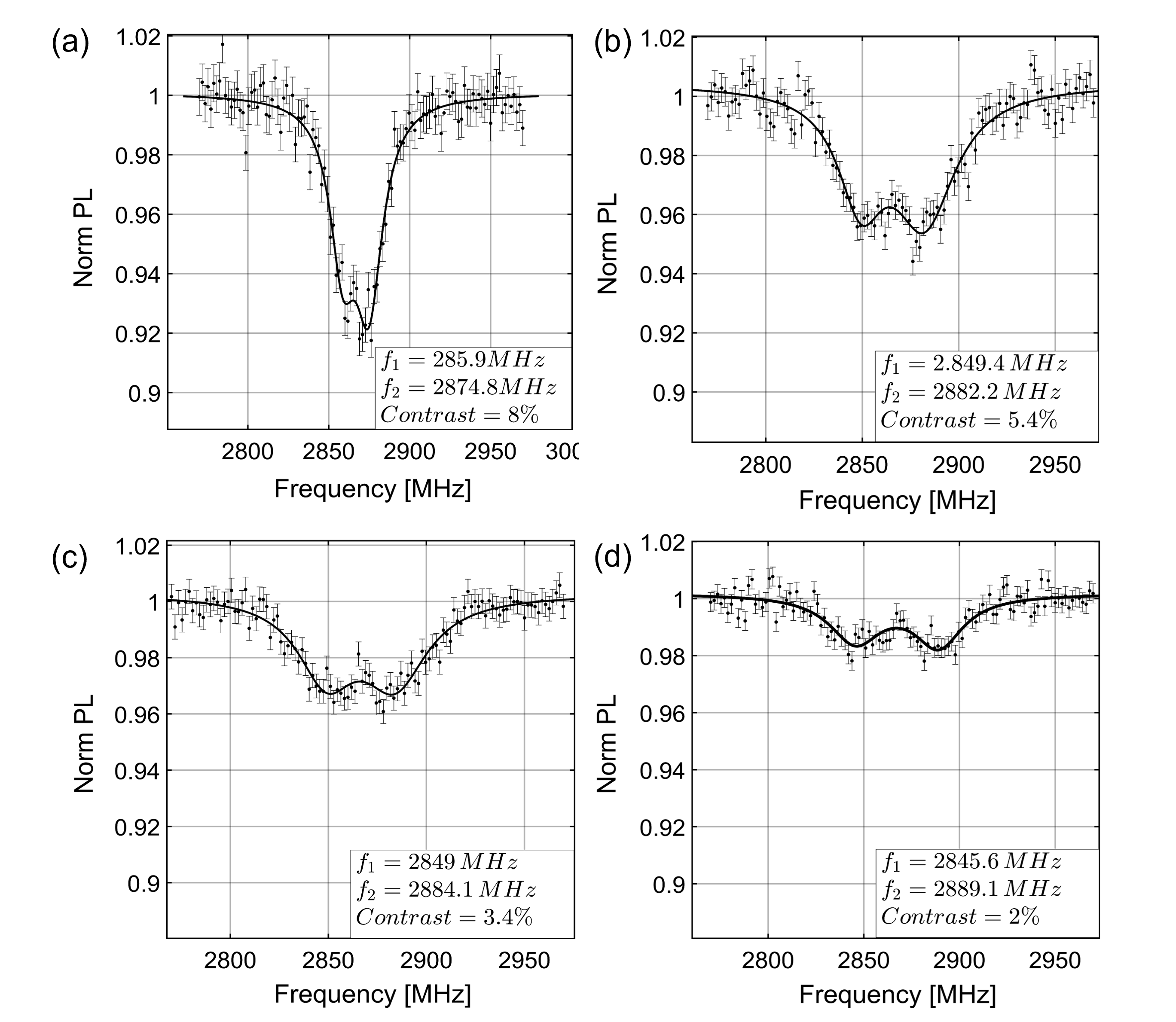}
\caption{\textbf{Continuous-wave ODMR spectra of trapped vaterite--40-nm-nanodiamond hybrids at applied fields of 0, 0.58, 0.63 and 0.77 mT (0, 5.8, 6.3 and 7.7 G).}}
\phantomsection\label{fig:C1}
\end{figure}
\\Having established that the assembled and trapped material retains a field-dependent ODMR response, the remaining sections address the second sensing mode: how the hydrated carrier interface changes the NV longitudinal-relaxation rate.

\subsection*{C2. Effective model for proton-responsive relaxometry: Physical interface}
\renewcommand{\thefigure}{C2.\arabic{figure}}
\setcounter{figure}{0}
The effective model represents a layered material junction.The vaterite contributes carbonate- and calcium-associated sites and a porous hydrated surface; PSS contributes fixed sulfonate charges, counterions and a polymer-mediated separation between solids; the nanodiamond contributes oxygenated surface groups, electronic defects and near-surface NV centres; and the liquid contributes solvent structure, dissolved ions, buffer species and proton activity. These components occupy the same interfacial region but are not resolved separately by the present measurements.

The surface-bound geometry is what makes this coarse-grained description physically relevant.The NV response is weighted towards nearby fluctuators, while the carrier porosity and polymer layer determine how the liquid reaches and reorganises that local environment. The model is therefore a bridge from bulk dosing to the NV observable, not an atomistic claim that a single protonation site or a single field channel has been identified.

\subsubsection*{C2.1. Scope of the model}
\renewcommand{\thefigure}{C2.1.\arabic{figure}}
\setcounter{figure}{0}
This note expands the grand-canonical charge-regulation and
spectral-transfer framework introduced in main-text equations (3)--(6).
The control variable is measured bulk pH or nominal proton-equivalent
dose. The transduction sequence runs from that variable to local
activity, surface potential, effective occupation, chemical switching, a
faster relaxation-active spectrum and \ensuremath{\Gamma_{1,\mathrm{eff}}} \ensuremath{\equiv} 1/\ensuremath{T_1}. The model organises the trends while keeping
thermodynamics, kinetics and quantum readout distinct; it is not a
uniquely identified microscopic fit \cite{ref31}.

\begin{center}
\small\itshape
bulk pH or nominal H\textsuperscript{+} dose $\rightarrow$ local interfacial activity
$\rightarrow$ surface/site occupation $\rightarrow$ site switching and surface dynamics
$\rightarrow S(\omega) \rightarrow \Gamma_1 = 1/T_1$
\end{center}

Potential contributors include carbonate-like vaterite sites, hydrated
calcium sites, PSS sulfonate groups and counterions, adsorbed water,
nanodiamond surface groups and near-surface defects. The diagram
proceeds from the bulk medium at the left, through local surface
activity, an effective partition function and kinetic profile, to
interfacial field, charge or dipole noise and then NV \ensuremath{T_1} at the right.
These species are not fitted independently; they are grouped into
effective site classes so that the limited data do not imply an
unsupported microscopic assignment.


\subsubsection*{C2.2 NV relaxation observable}
The measured contrast is fitted with

\begin{equation}
C(\tau)=C_0+A\exp\!\left[-\left(\frac{\tau}{T_1}\right)^n\right].\label{eq:s7}
\end{equation}

where n captures the distribution of local NV environments within  the nanodiamond layer.The effective longitudinal-relaxation rate is

\begin{equation}
\Gamma_{1,\mathrm{eff}}\equiv\frac{1}{T_1}.\label{eq:s8}
\end{equation}

\emph{We write the measured rate as}

\begin{equation}
\Gamma_{1,\mathrm{eff}}(u)
=\Gamma_{1,\mathrm{eff},0}+\Delta\Gamma_{\mathrm{chem}}(u)+\Gamma_{\mathrm{opt}}.\label{eq:s9}
\end{equation}

\ensuremath{\Gamma_{1,\mathrm{eff},0}} is the reference inverse characteristic time, \ensuremath{\Delta}\ensuremath{\Gamma}chem is the medium-dependent interfacial contribution and \ensuremath{\Gamma}opt is the optical-control contribution. \ensuremath{\Gamma_{1,\mathrm{eff}}} is used as a comparison metric for ensemble stretched-exponential decays; microscopic rate additivity is exact only for exponential Markovian relaxation. The Fig. 5 controls show no resolved \ensuremath{T_1} change from the 976-nm trap over the tested power range, whereas 532-nm illumination during the dark interval adds a substantial optical response.
For the transverse magnetic channel written explicitly in main-text equation (\ref{eq:main-2}), the chemical increment is
\begin{equation}
\Delta\Gamma_1^{(\perp B)}(u)
=\gamma_e^2S_{B_{\perp}}^{\mathrm{chem}}(\omega_{\mathrm{NV}},u).
\label{eq:s10}
\end{equation}

The chemical transverse magnetic-noise spectrum in Supplementary Equation \ref{eq:s10} follows one angular-frequency PSD convention. Its normalisation absorbs ensemble geometry and orientation. Proton activity may regulate electronic surface-spin populations and correlation times without the proton nuclear moment supplying GHz noise. Charge, electric-field, strain and defect-state changes remain possible upstream regulators; Supplementary Equation (\ref{eq:s10}) therefore represents one microscopic channel, not a unique assignment of \ensuremath{\Gamma_{1,\mathrm{eff}}}.

\subsubsection*{C2.3 Bulk-to-surface activity}
For an aqueous reservoir, the proton chemical potential is
\begin{equation}
\mu_{\mathrm{H,bulk}}=\mu_{\mathrm H}^{0}
+k_{\mathrm B}T\ln a_{\mathrm{H,bulk}}.\label{eq:s11}
\end{equation}
Here the bulk proton activity is dimensionless on the selected standard state. At an interface with electrostatic potential \ensuremath{\psi_0}, electrochemical equilibrium gives
\begin{equation}
\mu_{\mathrm{H,loc}}=\mu_{\mathrm H}^{0}
+k_{\mathrm B}T\ln a_{\mathrm{H,bulk}}-e\psi_0.\label{eq:s12}
\end{equation}

\emph{or, equivalently,}

\begin{equation}
a_{\mathrm{H,loc}}=a_{\mathrm{H,bulk}}
\exp\!\left(-\frac{e\psi_0}{k_{\mathrm B}T}\right)
=a_{\mathrm{H,bulk}}\exp\!\left(-\frac{F\psi_0}{RT}\right).\label{eq:s13}
\end{equation}

\emph{and}

\begin{equation}
\mathrm{pH}_{\mathrm{loc}}=\mathrm{pH}_{\mathrm{bulk}}
+\frac{F\psi_0}{2.303RT}.\label{eq:s14}
\end{equation}

For \ensuremath{\psi_0} \textless{} 0, the Boltzmann factor enriches positive charge near the interface and the local pH is lower than the bulk value.The sign of the pH shift is therefore fixed by the electrostatic convention used in the preceding equations. The magnitude is not calculated here because \ensuremath{\psi_0} was not measured independently and would itself depend on vaterite, PSS, nanodiamond coverage, ionic strength and the adsorbed molecular layer.

In the DMEM and ethanol experiments, the control variable is nominal proton-equivalent dose rather than an independently measured interfacial proton activity. We therefore write

\begin{equation}
a_{\mathrm{H,loc}}=g_m(H_{\mathrm{nom}},B_m,I_m,\psi_0).\label{eq:s15}
\end{equation}

The medium-dependent mapping in Supplementary Equation (S15) includes buffer capacity, ionic strength, solvent activity, surface potential, adsorption and accessibility of the porous/polymer-coated interface. Its explicit form is not inferred from the limited dose series. This distinction is essential for DMEM and ethanol: the nominal amount added to the cuvette is an experimental coordinate, not a direct measurement of free proton activity at an NV centre.
\subsubsection*{C2.4 Carbonate coordinate and its relation to acidity constants}
\renewcommand{\thefigure}{C2.4.\arabic{figure}}
\setcounter{figure}{0}
\emph{The aqueous carbonate pair is described by}

\begin{equation}
\mathrm{CO_2^{*}}+\mathrm{H_2O}
\rightleftharpoons\mathrm{H^+}+\mathrm{HCO_3^-}.\label{eq:s16}
\end{equation}

\emph{and}

\begin{equation}
\mathrm{HCO_3^-}\rightleftharpoons\mathrm{H^+}+\mathrm{CO_3^{2-}}.\label{eq:s17}
\end{equation}

\emph{For the second equilibrium,}

\begin{equation}
K_{a2}=\frac{a_{\mathrm{H^+}}a_{\mathrm{CO_3^{2-}}}}
{a_{\mathrm{HCO_3^-}}}.\label{eq:s18}
\end{equation}

\emph{Using concentration ratios as an effective coordinate,}

\begin{equation}
\frac{[\mathrm{HCO_3^-}]}{[\mathrm{CO_3^{2-}}]}
=10^{\mathrm pK_{a2}-\mathrm{pH}_{\mathrm{bulk}}}.\label{eq:s19}
\end{equation}

We use \ensuremath{\mathrm{p}K_{a2}} \ensuremath{\approx} 10.33 only as an aqueous bicarbonate--carbonate coordinate. It is not assigned as the intrinsic acidity constant of a specific vaterite, PSS or nanodiamond surface site, and it is not
applied quantitatively to ethanol.

\begin{figure}[!htbp]
\centering
\includegraphics[width=0.8\linewidth,height=0.70\textheight,keepaspectratio]{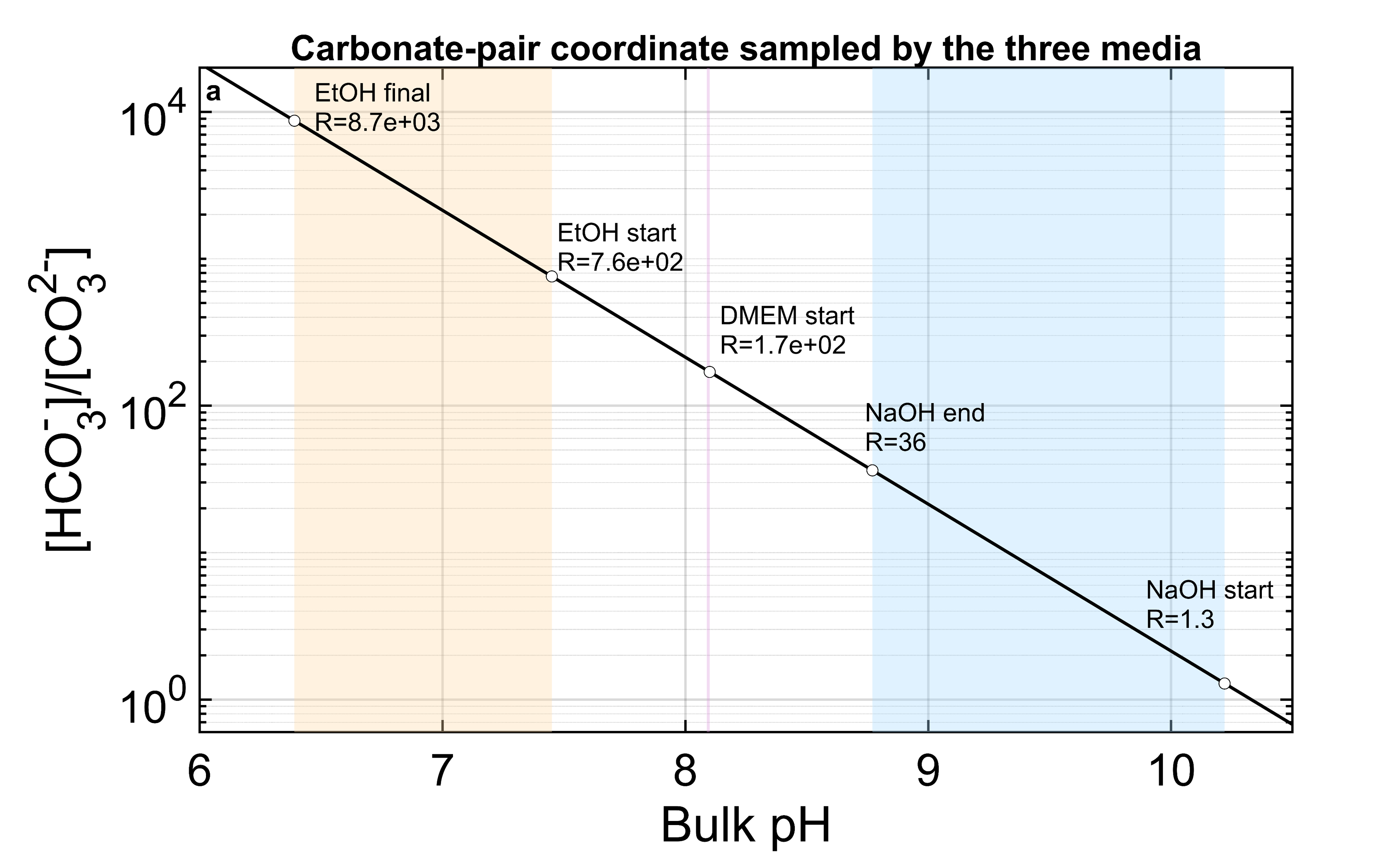}
\caption{\textbf{Overview of the carbonate-pair coordinate on a bulk-or-apparent-pH axis, including the NaOH/HCl, DMEM and ethanol measurement windows.}}
\phantomsection\label{fig:C2.2a}
\end{figure}

 Fig. \ref{fig:C2.2a} uses a bulk-or-apparent-pH axis to place the NaOH/HCl, DMEM and ethanol conditions on a common visual coordinate. The NaOH/HCl series spans the low-ratio to higher-ratio portion of the aqueous carbonate-pair curve, while DMEM occupies a narrower bicarbonate-dominated interval. Ethanol measurements are retained only as apparent-pH comparisons and are not converted to an aqueous thermodynamic pH scale.
\subsubsection*{C2.5 Effective surface-site occupation}
\renewcommand{\thefigure}{C2.5.\arabic{figure}}
\setcounter{figure}{0}
For each effective site class s, minimising main-text equation (\ref{eq:main-5}) at
fixed \ensuremath{\psi_0} gives the two-state occupation below. This is the conditional independent-site limit. In the full charge-regulated interface, protonation changes \ensuremath{\sigma} and \ensuremath{\psi_0} and the equilibrium covariance follows the inverse Hessian of \ensuremath{\Omega}.

\emph{The occupation is represented by}

\begin{equation}
f_s(\psi_0)=\frac{1}{1+10^{\mathrm{pH}_{\mathrm{loc}}-\mathrm pK_{a,s}^{\mathrm{eff}}}}.\label{eq:s20}
\end{equation}

For \ensuremath{N_s} independent effective sites at fixed \ensuremath{\psi_0}, the conditional variance
is the first line of Supplementary Equation (S21); the second line
states the full covariance form including electrostatic feedback.

\begin{equation}
\begin{aligned}
\operatorname{Var}(N_{\mathrm H,s}\mid\psi_0)&=N_s f_s(1-f_s),\\
\operatorname{Cov}(\mathbf N)&=k_{\mathrm B}T
\left[\nabla_{\mathbf N}^{2}\Omega\right]^{-1}.
\end{aligned}\label{eq:s21}
\end{equation}

The conditional variance is maximal at half occupation, where the local pH equals the effective site-acidity coordinate.

\begin{equation}
f_s=\frac{1}{2}\quad\Rightarrow\quad
\mathrm{pH}_{\mathrm{loc}}\approx\mathrm pK_{a,s}^{\mathrm{eff}}.\label{eq:s22}
\end{equation}

\begin{figure}[!htbp]
\centering
\includegraphics[width=\linewidth,height=0.80\textheight,keepaspectratio]{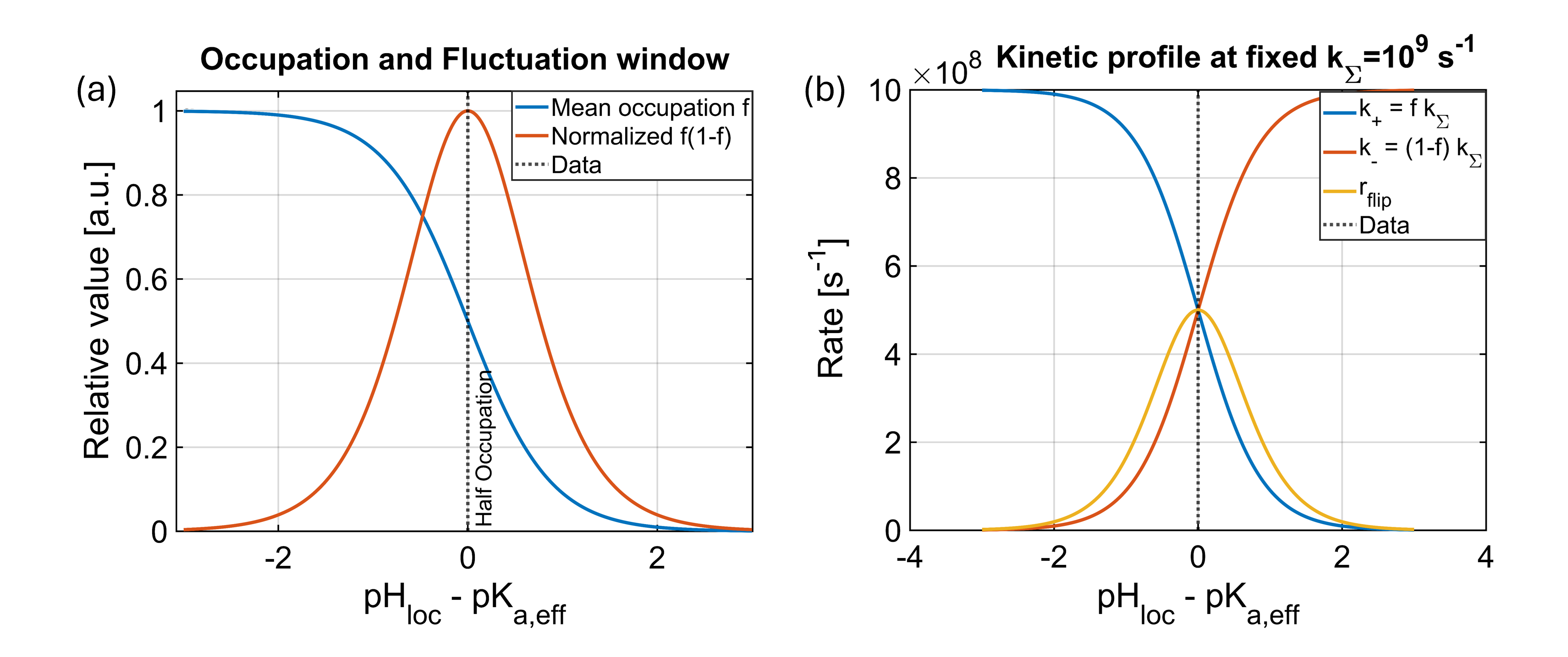}
\caption{\textbf{Effective surface-site occupation and switching profile.}(a) mean occupation and binomial variance versus the local-pH offset from the effective acidity coordinate.(b) Illustrative forward and reverse rates and switching flux for a fixed total rate of 10\textsuperscript{9} s\ensuremath{^{-1}}}
\phantomsection\label{fig:C2.5}
\end{figure}

We use the binomial factor in Supplementary Equation (\ref{eq:s21}) as an effective fluctuation amplitude because a two-state population has little switching variance when almost every site is empty or occupied and the largest variance near half occupation. Supplementary Fig. \ref{fig:C2.5} shows this sequence directly: the left panel compares the monotonic mean
occupation with the peaked variance, and the right panel shows the
corresponding illustrative forward rate, reverse rate and switching flux
for a fixed total rate. This construction does not identify a unique
chemical site or prove that site occupation is the only changing
material variable.

\subsubsection*{C2.6 Effective switching kinetics}

To connect equilibrium occupation to a noise spectrum, each
effective site class is represented as a two-state process,

\begin{equation}
S_s+\mathrm H_{\mathrm{loc}}^+\rightleftharpoons S_s\mathrm H.\label{eq:s23}
\end{equation}

with dynamics

\begin{equation}
\frac{df_s}{dt}=k_{+,s}(1-f_s)-k_{-,s}f_s.\label{eq:s24}
\end{equation}

\emph{The equilibrium occupation and correlation time are}

\begin{equation}
f_{s,\mathrm{eq}}=\frac{k_{+,s}}{k_{+,s}+k_{-,s}},
\qquad \tau_s=\frac{1}{k_{+,s}+k_{-,s}}.\label{eq:s25}
\end{equation}

Detailed balance combines a pseudo-first-order association rate
proportional to local proton activity with the reverse rate and gives

\begin{equation}
k_{+,s}=k_{\mathrm{on},s}a_{\mathrm{H,loc}},
\qquad \frac{k_{+,s}}{k_{-,s}}
=\frac{a_{\mathrm{H,loc}}}{K_{a,s}^{\mathrm{eff}}}
=10^{\mathrm pK_{a,s}^{\mathrm{eff}}-\mathrm{pH}_{\mathrm{loc}}}.\label{eq:s26}
\end{equation}

Writing the total switching rate as the sum of the forward and reverse
rates,

\begin{equation}
k_{+,s}=f_s k_{\Sigma,s},
\qquad k_{-,s}=(1-f_s)k_{\Sigma,s}.\label{eq:s27}
\end{equation}

the per-site switching flux is

\begin{equation}
r_{\mathrm{flip},s}=(1-f_s)k_{+,s}+f_s k_{-,s}
=2f_s(1-f_s)k_{\Sigma,s}.\label{eq:s28}
\end{equation}

This flux is largest near half occupation if the total switching
rate is not itself suppressed.

After a step in activity, the occupation approaches its new
equilibrium as

\begin{equation}
f_s(t)=f_{s,\mathrm{eq}}^{\mathrm{new}}
+\left[f_s(0)-f_{s,\mathrm{eq}}^{\mathrm{new}}\right]
\exp\!\left(-\frac{t}{\tau_s}\right).\label{eq:s29}
\end{equation}

The fitted data do not independently determine access, hydration and
exchange contributions to \ensuremath{\tau_{\mathrm{chem}}}. We retain \ensuremath{\tau_{\mathrm{chem}}} as an effective
chemical correlation time and keep it distinct from the faster electronic or defect-state time \ensuremath{\tau_e} in main-text equation \ref{eq:main-6}.
Any microscopic timescale quoted below is a spectral constraint and must not be interpreted as a measured proton exchange rate.
The kinetic description adds the missing time dimension to the equilibrium occupation model. A site can have a large thermodynamic variance but contribute little at the NV transition if it switches too slowly or too quickly for the selected spectral window. The next subsection therefore converts the same two-state process into a frequency-dependent noise spectrum.
\begin{table}[htbp]
\centering
\caption{Definitions of parameters used in the effective two-state protonation kinetics.}
\label{tab:kinetic_parameters}
\small
\begin{tabular}{|l|l|l|l|}
\hline
Symbol & Definition & Units & Physical interpretation \\
\hline
$S_s$ & Unoccupied effective site & -- & Deprotonated thermodynamic state \\
\hline
$S_s\mathrm{H}$ & Occupied effective site & -- & Protonated thermodynamic state \\
\hline
$f_s$ & Occupation probability & -- & Probability that the site is protonated \\
\hline
$1-f_s$ & Vacancy probability & -- & Probability that the site is unoccupied \\
\hline
$k_{+,s}$ & Forward switching rate & $\mathrm{s}^{-1}$ & Effective proton-capture rate from the local reservoir \\
\hline
$k_{-,s}$ & Reverse switching rate & $\mathrm{s}^{-1}$ & Effective deprotonation/back-transfer rate \\
\hline
$k_{\mathrm{on},s}$ & Proton association rate coefficient & $\mathrm{s}^{-1}$ & Rate coefficient for proton capture; $k_{+,s}=k_{\mathrm{on},s}a_{\mathrm{H,loc}}$ \\
\hline
$k_{\Sigma,s}$ & Total switching rate & $\mathrm{s}^{-1}$ & Sum of forward and reverse rates; $k_{\Sigma,s}=k_{+,s}+k_{-,s}$ \\
\hline
$r_{\mathrm{flip},s}$ & Per-site switching flux & $\mathrm{s}^{-1}$ & Average rate of protonation/deprotonation switching events \\
\hline
$\tau_s$ & Chemical correlation time & $\mathrm{s}$ & Characteristic relaxation time; $\tau_s=1/k_{\Sigma,s}$ \\
\hline
$a_{\mathrm{H,loc}}$ & Local proton activity & -- & Local thermodynamic driving force for protonation \\
\hline
$K_{a,s}^{\mathrm{eff}}$ & Effective acid dissociation constant & -- & Effective equilibrium constant for protonation/deprotonation \\
\hline
$\mathrm{p}K_{a,s}^{\mathrm{eff}}$ & Effective acid dissociation parameter & -- & Equilibrium parameter controlling the protonation state \\
\hline
\end{tabular}
\end{table}

\subsubsection*{C2.7 Noise spectrum and frequency matching}
\renewcommand{\thefigure}{C2.7.\arabic{figure}}
\setcounter{figure}{0}
For a two-state Markov process used as the minimal occupation-switching
closure,

\begin{equation}
\left\langle\delta n_s(t)\delta n_s(0)\right\rangle
=f_s(1-f_s)\exp\!\left(-\frac{|t|}{\tau_{\mathrm{chem},s}}\right).\label{eq:s30}
\end{equation}

The occupation-noise spectrum is

\begin{equation}
S_{n_s}(\omega)=\frac{2f_s(1-f_s)\tau_{\mathrm{chem},s}}
{1+\omega^2\tau_{\mathrm{chem},s}^{2}}.\label{eq:s31}
\end{equation}

If chemical switching itself changes a transverse relaxation-active
field, its Lorentzian contribution at angular frequency \ensuremath{\omega} is maximal
when

\begin{equation}
\omega\tau_{\mathrm{chem},s}=1.\label{eq:s32}
\end{equation}

The matching time is therefore

\begin{equation}
\tau_{\mathrm{match}}=\frac{1}{\omega}=\frac{1}{2\pi f}.\label{eq:s33}
\end{equation}

For the 2.87-GHz NV single-quantum transition,

\begin{equation}
\tau_{\mathrm{match}}(2.87\,\mathrm{GHz})\approx55\,\mathrm{ps}.\label{eq:s34}
\end{equation}

At the 0.5--0.8-mT fields used for ODMR, the proton Larmor
frequency is

\begin{equation}
f_p=\frac{\gamma_p}{2\pi}B
\approx42.58\,\mathrm{MHz\,T^{-1}}\times B.\label{eq:s35}
\end{equation}

which gives

\begin{equation}
B=0.5-0.8\,\mathrm{mT}\quad\Rightarrow\quad
f_p\approx21-34\,\mathrm{kHz}.\label{eq:s36}
\end{equation}

Direct proton nuclear-spin precession is far from the 2.87-GHz
single-quantum transition at the measured fields. The 55-ps condition
constrains a downstream fast electronic, paramagnetic or defect-state
fluctuator; it is not assigned to ordinary protonation. Slow chemical
occupation can instead regulate the population, coupling or \ensuremath{\tau_e} of that
faster bath.

\begin{figure}[!t]
\centering
\includegraphics[width=0.7\linewidth,height=0.60\textheight,keepaspectratio]{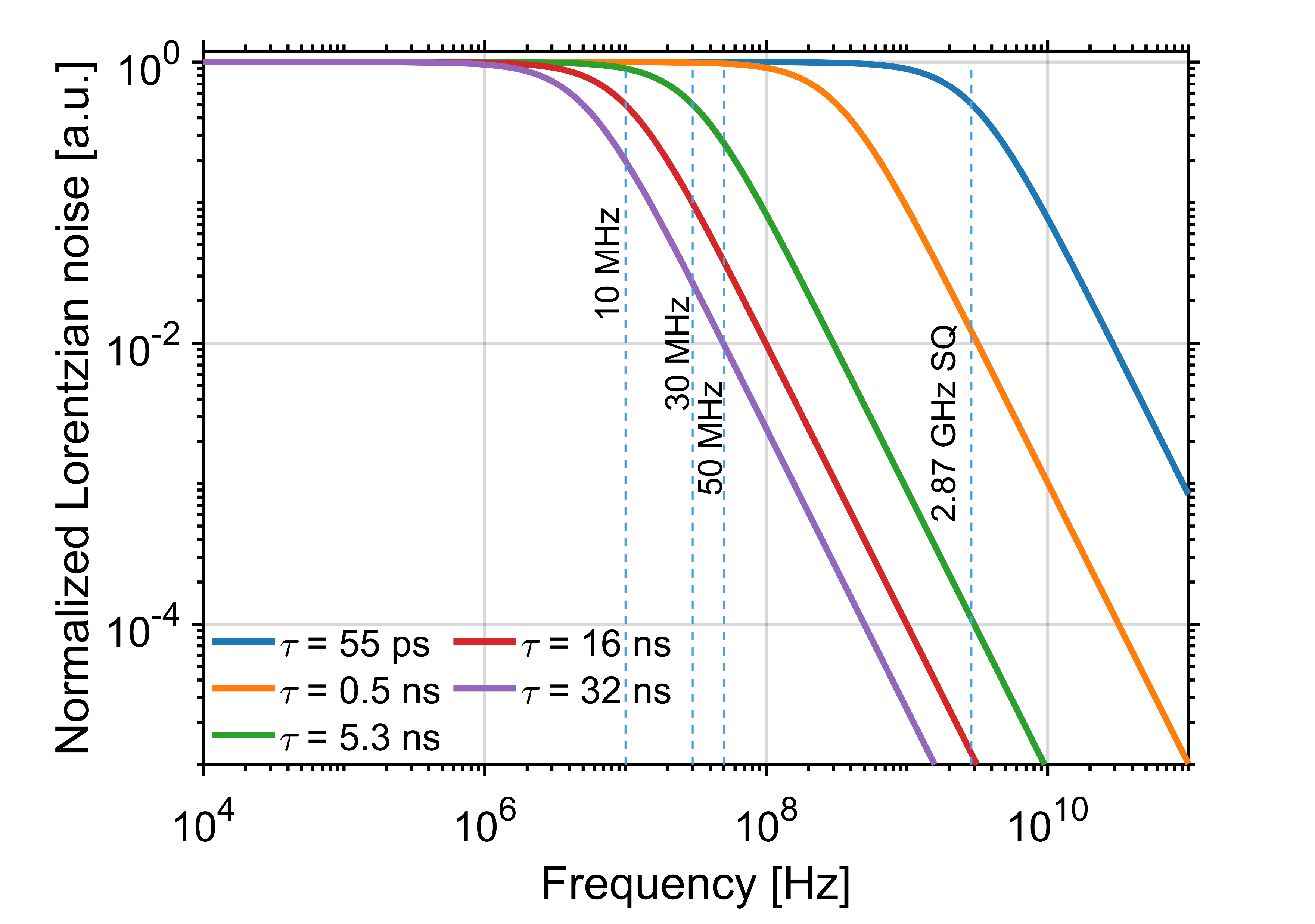}
\caption{\textbf{Frequency selection by NV relaxometry.} Lorentzian spectra are shown for representative correlation times. The 2.87-GHz single-quantum transition is maximally weighted by a sub-100-ps downstream relaxation-active fluctuator, whereas MHz-scale channels sample slower dynamics. The times are illustrative inputs, not fitted proton-exchange times.}
\phantomsection\label{fig:C2.7.1}
\end{figure}

 Fig. \ref{fig:C2.7.1} visualises the formal Lorentzian frequency
selection. The curves show representative correlation times and the
vertical markers identify MHz-scale channels and the 2.87-GHz
transition. Their role is to illustrate spectral matching, not to report
measured protonation kinetics.

\subsubsection*{C2.8 Coupling from surface dynamics to the NV layer}
The chemical contribution is represented generically as

\begin{equation}
\begin{aligned}
\Delta\Gamma_{1,\mathrm{eff},m}(u)
={}&\sum_sN_{s,\mathrm{eff}}G_s\Bigl[
a_{s,m}f_s(u)L(\omega_0,\tau_{\mathrm e,s,m})\\
&\qquad{}+b_{s,m}f_s(u)(1-f_s(u))
L(\omega_0,\tau_{\mathrm{chem},s,m})\Bigr].
\end{aligned}
\label{eq:s37}
\end{equation}

\ensuremath{N_{s,\mathrm{eff}}} is the effective number of sites sampled by the NV near field, \ensuremath{G_s}
is a geometric and microscopic coupling factor, \ensuremath{f_s} describes occupation
gating, and \ensuremath{f_s}(1 \ensuremath{-} \ensuremath{f_s}) is the fixed-potential fluctuation weight.
Equation (\ref{eq:s37}) retains separate \ensuremath{\tau_e} and \ensuremath{\tau_{\mathrm{chem}}} timescales,
matching the structure of main-text equation (\ref{eq:main-6}). Neither timescale is
extracted uniquely from the present dose series.

The DMEM--ethanol contrast is therefore attributed qualitatively
to the combined effects of hydration, buffer capacity, proton activity,
exchange kinetics, ionic screening, adsorption, PSS counterions and
nanodiamond surface-defect dynamics.No single bulk material parameter
is sufficient to predict the response.In particular, a bulk
dielectric-loss expression is not interpreted as a first-principles
transmission coefficient for all magnetic, electric and defect-mediated
interfacial noise.

\subsubsection*{C2.9 Empirical fitting form for DMEM and ethanol}

The DMEM and ethanol datasets contain few dose points and use nominal
dose u rather than measured local activity. We therefore use the
restricted empirical Hill-like projection

\begin{equation}
T_1(u)=\frac{T_0}{1+(u/u_{50})^m}.\label{eq:s38}
\end{equation}

with \ensuremath{T_0} fixed by the zero-dose value and \ensuremath{u_{50}} an empirical scale in the
same nominal-dose units as u. This form is used only over the measured
interval; its \ensuremath{T_1} \ensuremath{\rightarrow} 0 high-dose limit is not extrapolated. The
corresponding effective inverse time is

\begin{equation}
\Gamma_{1,\mathrm{eff}}(u)=\Gamma_0\left[1+(u/u_{50})^m\right],
\qquad \Gamma_0=T_0^{-1}.\label{eq:s39}
\end{equation}

The local slope underlying the source-reported dose-response metric is
\begin{equation}
\frac{dT_1}{du}=-\frac{T_0m}{u_{50}}
\frac{(u/u_{50})^{m-1}}{\left[1+(u/u_{50})^m\right]^2}.\label{eq:s40}
\end{equation}
\ensuremath{u_{50}} is an empirical response scale, not a \ensuremath{\mathrm{p}K_a}, free proton concentration
or microscopic binding constant; m is a shape parameter, not evidence of
molecular cooperativity.Table \ref{tab:C2.1} reports the descriptive fit
parameters. Table \ref{tab:C2.2} retains the source-reported response metrics for
traceability; they are not independently noise-normalised because
acquisition time, measurement variance and particle-to-particle
dispersion are unavailable. The three-point ethanol fit has zero
residual degrees of freedom after fixing \ensuremath{T_0} and fitting \ensuremath{u_{50}} and m.
\\

\begin{table}[htbp]
\centering
\caption{Restricted empirical fit parameters for the DMEM and ethanol nominal-dose series.}
\label{tab:C2.1}
\small
\begin{tabular}{|l|l|l|l|}
\hline
Medium & $u_{50}$ ($\mu$M nominal dose) & $m$ (shape) & $\chi^2$ (source; weighting/df unavailable) \\
\hline
DMEM & 10.0 & 6.91 & 0.550 \\
\hline
Ethanol & $6.23 \times 10^3$ & 3.13 & not determined \\
\hline
\end{tabular}
\end{table}
\begin{table}[htbp]
\centering
\caption{Source-reported fit-derived response metrics retained for traceability; independent noise normalisation is unavailable.}
\label{tab:C2.2}
\small
\begin{tabular}{|l|l|l|}
\hline
Medium & Source-reported pH-equivalent response metric & Source-reported nominal-dose response metric \\
\hline
DMEM & $6.54 \pm 0.02$ mpH Hz$^{-1/2}$ & $6.46 \pm 0.04\,\mu$M Hz$^{-1/2}$ \\
\hline
Ethanol & $0.664 \pm 0.0034$ pH Hz$^{-1/2}$ & $6.33 \times 10^3 \pm 38\,\mu$M Hz$^{-1/2}$ \\
\hline
\end{tabular}
\end{table}
The approximately 623-fold separation between the fitted \ensuremath{u_{50}} values
describes a sharper response per nominal amount in DMEM. It does not establish a 623-fold difference in free interfacial proton concentration because the two media differ simultaneously in buffering, ionisation, hydration and adsorption.
\subsubsection*{C2.10 Regional description of the NaOH/HCl series}

The NaOH/HCl series spans pH 10.22--8.77 and is summarised by regional
fits because the measured slope changes across the sampled alkaline
range. Unlike the nominal-dose series, this experiment records a broad
aqueous pH change and crosses the carbonate-pair coordinate in
Supplementary Fig. C2.2. The empirical effective-rate form is

\begin{equation}
\Gamma_{1,\mathrm{eff}}(u)=\Gamma_0+S\,C_{\mu}(u;K,m).\label{eq:s41}
\end{equation}

\emph{with}

\begin{equation}
f(u;K,m)=\frac{(u/K)^m}{1+(u/K)^m},
\qquad C_{\mu}(u;K,m)=f(u;K,m)[1-f(u;K,m)].\label{eq:s42}
\end{equation}

Here u is the nominal acid-dose coordinate used in the source fit, K has
the same units and C\ensuremath{\mu} is defined in  Equation (S42).The
regional parameters are listed in Table \ref{tab:C2.3} in the order of the pH
windows:

\begin{table}[htbp] \centering \caption{Regional empirical parameters for the NaOH/HCl series.} \label{tab:C2.3} \begin{tabular}{|l|l|l|} \hline Parameter & Region 1 & Region 2 \\ \hline pH window & 10.22--9.86 & 9.79--8.77 \\ \hline $\Gamma_0$ ($\mu$s$^{-1}$) & $0.0131 \pm 0.0003$ & $0.0194 \pm 0.0003$ \\ \hline $T_0$ ($\mu$s) & 76.5 & 49.5 \\ \hline $S$ ($\mu$s$^{-1}$) & $0.0267 \pm 0.002$ & $0.058 \pm 0.009$ \\ \hline $K$ ($\mu$M) & $186.3 \pm 10.1$ & $515 \pm 49.6$ \\ \hline $m$ (shape) & $5.4 \pm 1.23$ & $6.86 \pm 2.16$ \\ \hline Source-reported pH response metric & $0.25 \pm 0.0008$ pH Hz$^{-1/2}$ & $1.25 \pm 0.006$ pH Hz$^{-1/2}$ \\ \hline \end{tabular} \end{table}
The two parameter sets describe the data locally. Region 1 is pH 10.22--9.86 and Region 2 is pH 9.79--8.77, matching Fig. \ref{fig:6} a. Changes in \ensuremath{\Gamma_0}, S, K and m describe the slope changes but do not prove a phase transition, site identity or mechanistic crossover without additional points and formal model comparison.

\subsubsection*{C2.11 Interpretation across media}
Three coupled material factors organise the comparison: the occupation state set by local activity, the effective switching time that determines spectral overlap, and the overall coupling between the hydrated interface and the NV layer. The same nominal proton addition can therefore produce different \ensuremath{T_1} changes when any of these factors change with the medium.

NaOH/HCl scans an aqueous carbonate-active window and changes both
bulk pH and carbonate speciation. DMEM is a multicomponent aqueous,
bicarbonate-buffered medium in which the measured bulk pH remains nearly
constant while the nominal addition changes buffer composition, ionic
conditions and the local interface. Ethanol provides a less strongly
hydrated and less ionised comparison in which electrode readings are
only apparent values. These categories are qualitative material states;
they are not direct measurements of an open or closed dielectric gate.

The model therefore links bulk chemistry to local surface
activity, effective site statistics, switching kinetics, interfacial
noise and \ensuremath{T_1} without asserting a unique microscopic pathway. Its main
physical conclusion is modest but useful: the vaterite/PSS/nanodiamond
interface and the surrounding medium jointly set the noise spectrum, so
the carrier cannot be treated as a passive support when chemical
relaxometry is interpreted.

\subsubsection*{C2.12 Hydration control} 
\renewcommand{\thefigure}{C2.12.\arabic{figure}}
\setcounter{figure}{0}
Wetting 40-nm nanodiamonds on a coplanar waveguide with deionised water
shortened \ensuremath{T_1} from 91 \ensuremath{\pm} 17 \ensuremath{\mu\mathrm{s}} in air to 17.4 \ensuremath{\pm} 2.4 \ensuremath{\mu\mathrm{s}} (Fig.\ref{fig:C2.12.1}). The change shows that hydration strongly modifies the
near-surface relaxation environment. It does not, by itself, identify
hydrogen nuclei as the direct fluctuator because oxygen, ions,
charge-state stability and surface defects also change on wetting.
\begin{figure}[!htbp]
\centering
\includegraphics[width=\linewidth,height=0.80\textheight,keepaspectratio]{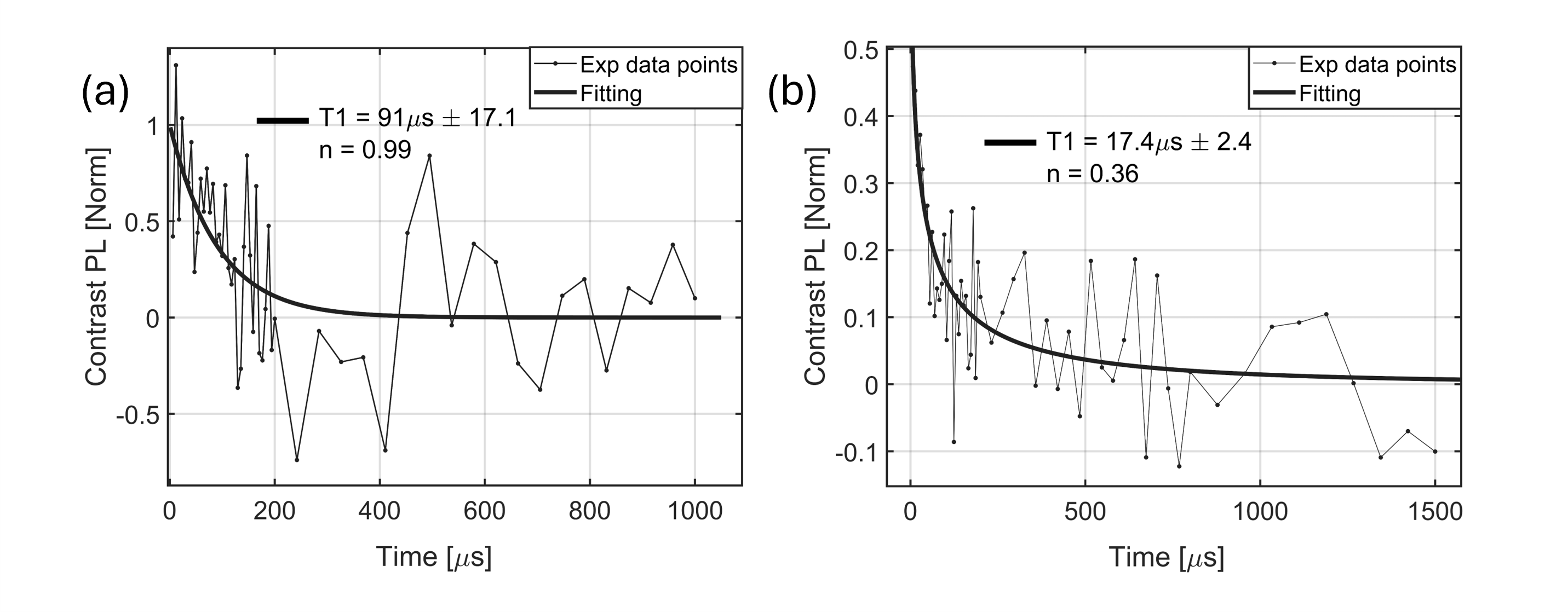}
\caption{\textbf{ \ensuremath{T_1} measurements for 40-nm nanodiamonds }(a) dried in air and (b) wetted with deionised water.}
\phantomsection\label{fig:C2.12.1}
\end{figure}
This hydration control provides a material bridge between the dry nanodiamond baseline and the liquid-phase hybrid measurements. Water changes the dielectric environment, surface termination, adsorbate mobility and access of dissolved species at the same time. The strong shortening therefore establishes sensitivity to the hydrated near-surface environment, while the more selective dose experiments are needed to separate that general hydration effect from acid-dependent changes.

\subsubsection*{C2.13 Additional dose-response data}
\renewcommand{\thefigure}{C2.13.\arabic{figure}}
\setcounter{figure}{0}

\paragraph{Ethanol and DMEM measurements}

\begin{figure}[H]
\centering
\includegraphics[width=\linewidth,height=0.45\textheight,keepaspectratio]{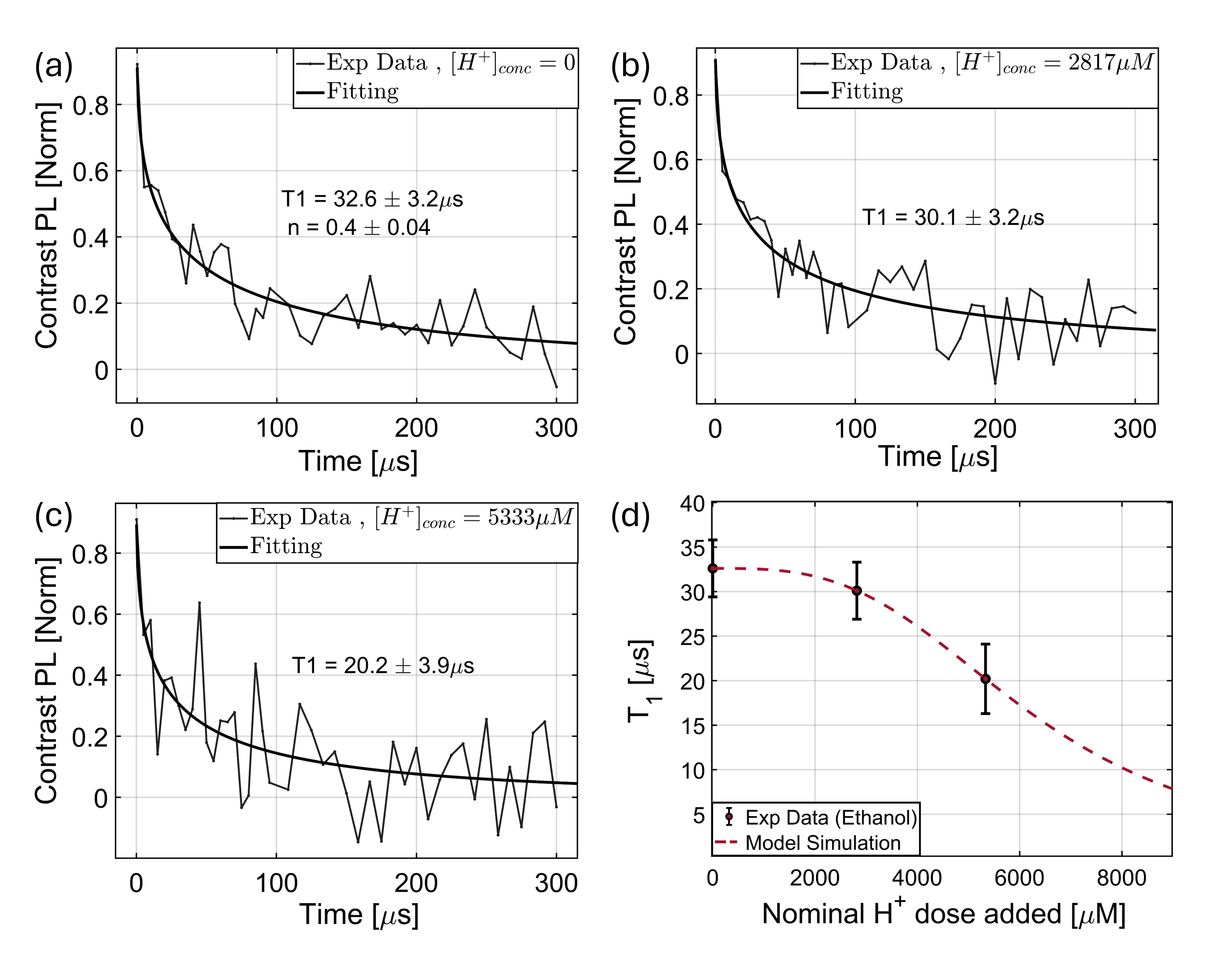}
\caption{\textbf{\ensuremath{T_1} decays for a trapped hybrid in ethanol at nominal additions of} (a) 0, (b) 2,817 and (c) 5,333 \ensuremath{\mu\mathrm{M}}. (d) Extracted \ensuremath{T_1} versus nominal dose with the empirical fit.}
\phantomsection\label{fig:figC2.13.1}
\end{figure}

\begin{figure}[ht]
\centering
\includegraphics[width=\linewidth,height=0.7\textheight,keepaspectratio]{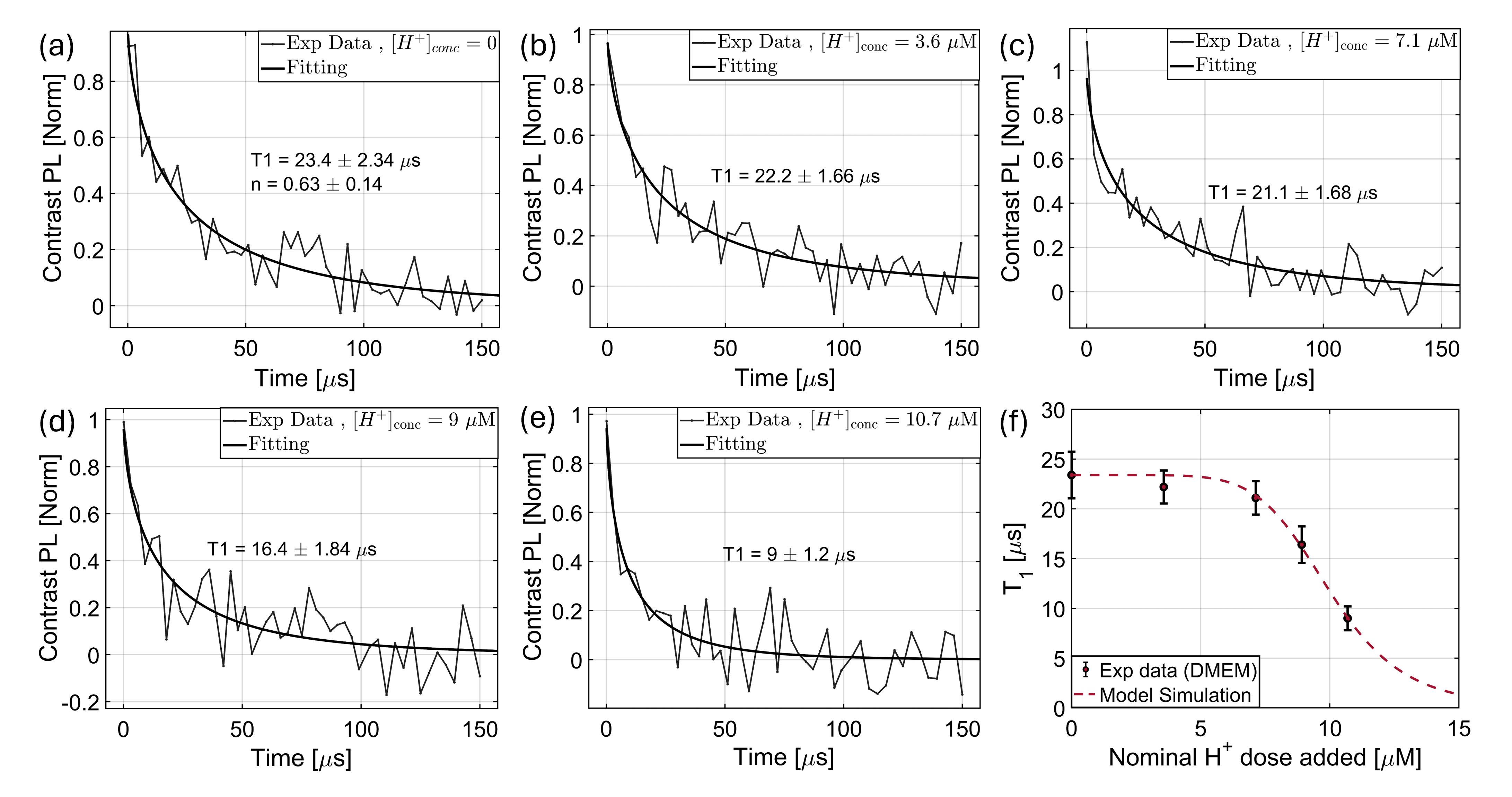}
\caption{\textbf{\ensuremath{T_1} decays for a trapped hybrid in ethanol at nominal additions of} (a) 0, (b) 2,817 and (c) 5,333 \ensuremath{\mu\mathrm{M}}. (d) Extracted \ensuremath{T_1} versus nominal dose with the empirical fit.}
\phantomsection\label{fig:figC2.13.2}
\end{figure}

The individual dose-response decays are presented in the same order as
their discussion: ethanol first in Fig. \ref{fig:figC2.13.1}, followed by
DMEM in Fig. \ref{fig:figC2.13.2}. Each figure shows the fitted decay
curves before the extracted \ensuremath{T_1}-versus-dose graph, allowing the reader to
trace every summary point back to the underlying time-domain data. The
mechanistic interpretation remains the effective model described in
C2.1–C2.11.

\paragraph{Dose-response curves}

In  Fig. \ref{fig:figC2.13.1}, panels a--c show the ethanol decays at
nominal additions of 0, 2,817 and 5,333 \ensuremath{\mu\mathrm{M}}. \ensuremath{T_1} changes from 32.6 \ensuremath{\pm} 3.2
\ensuremath{\mu\mathrm{s}} to 30.1 \ensuremath{\pm} 3.2 and 20.2 \ensuremath{\pm} 3.9 \ensuremath{\mu\mathrm{s}}; panel d collects the extracted
values and restricted empirical fit. The endpoint increment is
approximately 1.9 fit-standard-errors, and the three-point series does
not independently validate an isotherm or noise-normalised metric.
 Fig. \ref{fig:figC2.13.2} presents the DMEM series at nominal doses of 0,
3.6, 7.1, 9.0 and 10.7 \ensuremath{\mu\mathrm{M}}, with fitted \ensuremath{T_1} values of 23.4 \ensuremath{\pm} 2.3, 22.2 \ensuremath{\pm}
1.7, 21.1 \ensuremath{\pm} 1.7, 16.4 \ensuremath{\pm} 1.8 and 9.0 \ensuremath{\pm} 1.2 \ensuremath{\mu\mathrm{s}}. Panel f summarises the
trend and restricted fit. Buffer identity, potassium, phthalate, ionic
strength and solvent differ from the ethanol series, so the comparison
demonstrates medium dependence rather than H\ensuremath{^{+}} selectivity.
After the raw ethanol and DMEM decays, Table \ref{tab:C2.4} summarises the four
qualitative limits obtained by combining fast or slow spectral dynamics
with strong or weak effective coupling. The table is a regime map for
interpretation and is not fitted independently to the dose data.

\begin{table}[htbp]
\centering
\caption{Qualitative frequency-coupling regimes are used to organise the effective model.}
\label{tab:C2.4}

\begin{tabular}{|l|l|l|}
\hline
Regime & Condition & Interpretation \\
\hline
Fast, strongly coupled & 
$\omega_0\tau_{\mathrm{corr}} \lesssim 1$; effective coupling high & 
Response set mainly by fluctuation amplitude and kinetics \\
\hline
Fast, weakly coupled & 
$\omega_0\tau_{\mathrm{corr}} \lesssim 1$; effective coupling low & 
Coupling-limited response \\
\hline
Slow, strongly coupled & 
$\omega_0\tau_{\mathrm{corr}} \gg 1$; effective coupling high & 
High-frequency spectral weight is suppressed \\
\hline
Slow, weakly coupled & 
$\omega_0\tau_{\mathrm{corr}} \gg 1$; effective coupling low & 
Kinetics and coupling both suppress the response \\
\hline
\end{tabular}

\end{table}

\FloatBarrier\section*{C3. Reduced-point photoluminescence protocol}
\renewcommand{\thefigure}{C3.\arabic{figure}}
\setcounter{figure}{0}

\begin{figure}[!htbp]
\centering
\includegraphics[width=\linewidth,height=0.80\textheight,keepaspectratio]{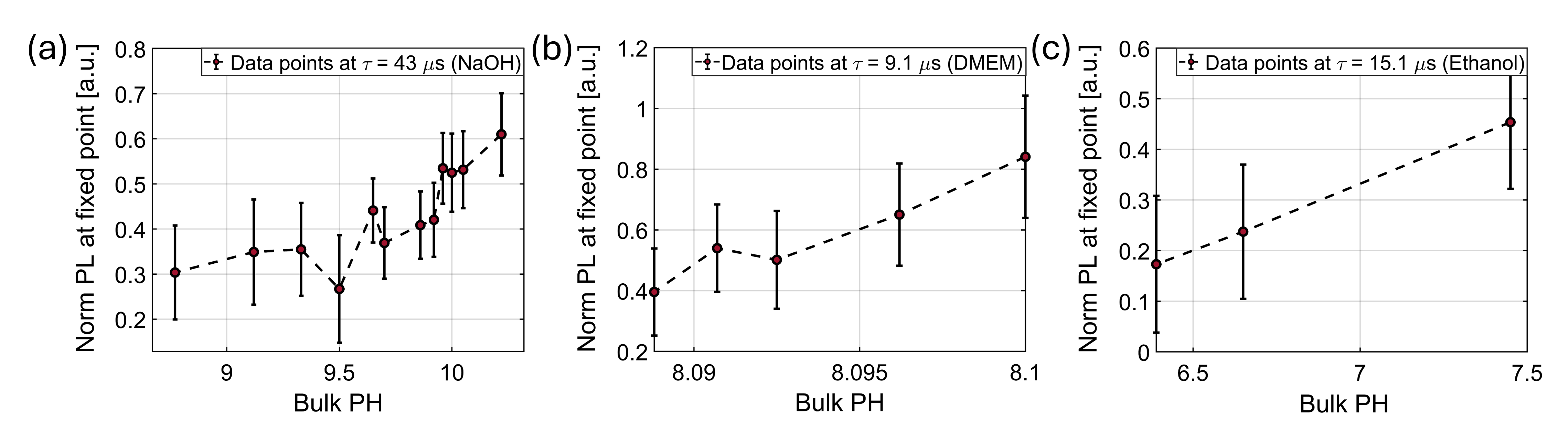}
\caption{\textbf{Reduced-point PL response at the selected delay for} (a) NaOH/HCl, (b) DMEM and (c) ethanol.}
\phantomsection\label{fig:C3.1}
\end{figure}

Full \ensuremath{T_1} relaxometry requires a complete decay curve at each chemical
condition. A faster protocol samples t = 0, a late reference and an
intermediate delay selected near the largest condition-dependent
contrast; temporal common-mode rejection offers a complementary route to
suppress technical noise \cite{ref32}.  Fig. \ref{fig:C3.1} presents the
response in the same environmental order used above: NaOH/HCl, DMEM and
ethanol. The intermediate signal is normalised to the measured extrema
and plotted against measured pH or nominal dose.

The reduced-point signal preserves the direction of the response while
shortening acquisition. The source analysis reported response metrics of
1 and 5.4 pH Hz\ensuremath{^{-1/2}} in the two NaOH/HCl regions, approximately 0.042
pH Hz\ensuremath{^{-1/2}} in DMEM and 6.99 pH Hz\ensuremath{^{-1/2}} in ethanol. These values are
protocol-specific and are not independently noise-normalised because
acquisition time, averaging and measurement variance are incomplete;
they should not be compared directly with the full-decay metrics.

